\documentclass[sigconf]{acmart}

\usepackage{tabularx}
\usepackage{subcaption}
\usepackage{todonotes}
\usepackage{booktabs}

\AtBeginDocument{%
  \providecommand\BibTeX{{%
    \normalfont B\kern-0.5em{\scshape i\kern-0.25em b}\kern-0.8em\TeX}}}

\copyrightyear{2020}
\acmYear{2020}
\setcopyright{rightsretained}

\acmConference{}
\acmDOI{}
\acmISBN{}
\acmBooktitle{}

\begin{document}

\title{Cryptoasset Competition and Market Concentration in the Presence of Network Effects}

\author{Konstantinos Stylianou}
\affiliation{%
 \institution{University of Leeds}
 \streetaddress{Woodhouse Lane}
 \city{Leeds}
 \country{United Kingdom}}
 \email{k.stylianou@leeds.ac.uk}

\author{Leonhard Spiegelberg}
\affiliation{%
  \institution{Brown University}
  \streetaddress{115 Waterman St}
  \city{Providence}
  \state{Rhode Island}
  \country{USA}}
  \email{lspiegel@cs.brown.edu}

\author{Maurice Herlihy}
\affiliation{%
  \institution{Brown University}
  \streetaddress{115 Waterman St}
  \city{Providence}
  \state{Rhode Island}
  \country{USA}}
  \email{herlihy@cs.brown.edu}

\author{Nic Carter}
\affiliation{%
  \institution{Coin Metrics}
  \streetaddress{}
  \city{Boston}
  \state{Massachusetts}
  \country{USA}}
  \email{nic@coinmetrics.io}

\begin{abstract}
When network products and services become more valuable as their userbase grows (network effects), this tendency can become a major determinant of how they compete with each other in the market and how the market is structured. Network effects are traditionally linked to high market concentration, early-mover advantages, and entry barriers, and in the cryptoasset market they have been used as a valuation tool too. The recent resurgence of Bitcoin has been partly attributed to network effects too. We study the existence of network effects in six cryptoassets from their inception to obtain a high-level overview of the application of network effects in the cryptoasset market. We show that contrary to the usual implications of network effects, they do not serve to concentrate the cryptoasset market, nor do they accord any one cryptoasset a definitive competitive advantage, nor are they consistent enough to be reliable valuation tools. Therefore, while network effects do occur in cryptoasset networks, they are not a defining feature of the cryptoasset market as a whole.
\end{abstract}


\maketitle

\section{Introduction}
The rapid appreciation and popularization of cryptoassets over the past few years has incited a large body of scholarship on understanding their behavior and their positioning in the market, particularly financial markets. As cryptoassets gradually became a household investment and transaction medium, they began to invite greater regulatory and investor scrutiny, which created the need to better understand their function as a market of their own and as market that forms part of the greater economy. While early analyses focused on simple economic illustrations of the functioning of cryptoasset networks in isolation, later work started exploring market-wide phenomena, including the dominance patterns of some cryptoassets over others. 

Since cryptoassets are based on blockchain networks and are therefore network markets, one important parameter that reflects and determines their behaviour is the relationship between their userbase and their value. This relationship has a long history in network markets under the theory of \textit{network effects}. Network effects theory states that the value of a product or service \(V\) is co-determined by its userbase \(u\). Then, for products or services that obey network effects, one can derive the value of the network, and therefore their relative value to each other too, for a given userbase assuming that the relationship between \(V\) and \(u\) is known, for example \(V \propto nlog(u)\), \(V \propto u^2\), \(V \propto 2^u\) etc. 

Initially, this insight attracted attention because of its predictive potential of cryptoasset valuation. Indeed a number of studies attempted to develop valuation models based on network effects that could be used by investors to predict the future value of their assets and the value of the market as a whole. However, the implications of network effects go far beyond valuation and, understood properly, they inform also the structure and competitiveness of the market making them a key input into policy-making and regulatory decisions. Most notably, markets that are characterized by network effects are commonly thought to be \textit{winner-take-all} markets, where first mover advantage is key, entry barriers are high, networks hit tipping points of no return, and contestable monopolies or high concentration can be the natural state of the market. This is for two reasons: firstly, because the value of joining a network is increasing in the number of other network adopters, because the bigger the number of existing adopters the greater the utility every new adopter derives from it (\textit{pure network effects}), and secondly, because for every new adopter joining the network, existing adopters also benefit (\textit{network externalities}). In both cases bigger equals better (everything else equal), creating  an incentive for users to join the network where the value will grow larger both for new and for existing users, which creates a snowball effect. This kind of power concentration in networks that exhibit network effects usually makes regulators uneasy, and therefore, if cryptoassets exhibit network effects, they would (and should) attract higher regulatory and investor scrutiny. 

Extant literature on network effects in cryptoassets is limited and has focused almost exclusively on confirming or rejecting, usually for Bitcoin only, a specific application of network effects, namely Metcalfe's law, which states that the value of a network is proportional to the square of its users (\(V \propto u^2\)), and, if confirmed, it would be a useful valuation tool. However, this line of literature presents only a binary distinction between the existence or not of a specific type of network effects, focuses only on valuation, uses sub-optimal data, and has also been temporally limited to the period before the recent resurgence in mid 2019, or excludes periods, therefore missing key parts in the cryptoasset market evolution. 

By contrast, our analysis takes a more comprehensive view of network effects in cryptoassets, and, while it confirms that network effects occur in cryptoassets, it shows that they do not have the usual implications associated with them in terms of according competitive advantages, resulting in market concentration, or serving as a reliable valuation tool. Firstly, we define network effects to occur when the value of a cryptoasset network changes supra- or infra-proportionately to changes in its userbase, thereby showing both positive and reverse network effects, while not being constrained by a specific version of network effects. We also use two proxies for value and userbase each to better capture what users perceive as the value of the network and how the network size (userbase) should be measured, and we base our results on cleaner vetted data. Moreover, we examine multiple cryptoassets to get a broader view of the industry, as opposed to previous works which focused on Bitcoin. Lastly, our analysis covers the entire history of the studied cryptoassets, which includes the valuation spikes and subsequent declines in 2014, 2017 and 2019. The spike in 2019 and the preceding decline from the heights of 2017 are particularly valuable because they help us show that the results obtained in previous studies which sampled only up to early 2018 do not hold based on more recent history. 

\begin{figure*}
        \centering
        \includegraphics[width=\textwidth]{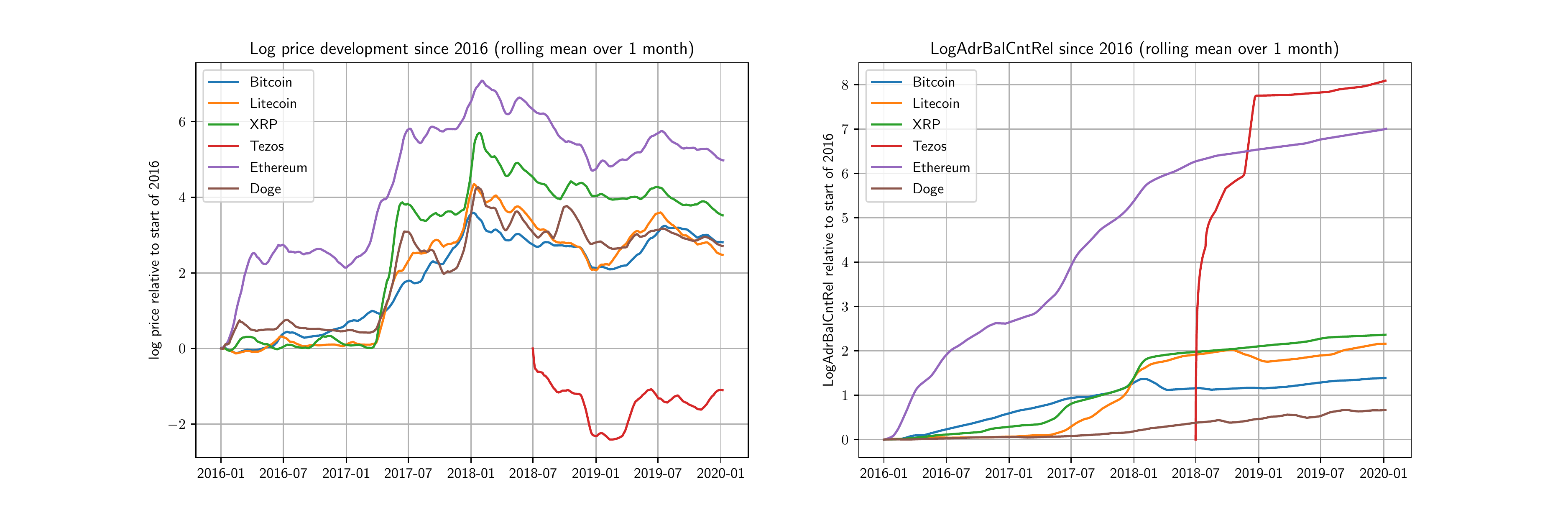}
         \caption{Price and userbase development since 2016.}
         \label{fig:logdev}
\end{figure*}

\section{Background, Motivation and Implications}
Network effects were first studied in the 1970s to more accurately capture the value and growth of telecommunications networks \cite{rohlfs_theory_1974}. The intuition was that when the nature of a product or service is such that it relies on linking users together, the value of the product \(V\) is co-determined by its userbase \(u\). More specifically, for every user added to the userbase of a product, value is created not just for the joining user but for existing users as well. As a result, each new user derives value from joining a network that is relative to the size of the network (pure network effects) and creates an externality in the form of value that is captured by the network of existing users (network externality). Conversely, for every exiting user, value is lost both for the exiting user and for existing users. This type of network effects was called direct network effects to distinguish it from later extensions to the theory, which accounted for the effects changes in the network's userbase have on complementary products and services developed for that network \cite{church_indirect_2008}. This latter type was called indirect network effects, and it is not the kind that will concern us here. 

The powerful implication of (direct) network effects is the increasing returns to the userbase and ultimately to the product exhibiting network effects. Because for products that exhibit network effects every new adopter makes the product more valuable relative to existing size of its network, it creates incentives for other adopters to adopt the product with the bigger network over its competitors. Consequently, the more the userbase grows the more it invites further growth rendering the product increasingly more valuable and competitive. The exact relationship between value and userbase can vary; While one can say that in the most basic version of network effects the value of a product grows linearly with the number of users added to its userbase (\(V \propto u\)) \cite{swann_functional_2002}, most commonly network effects are used to describe relationships that are logarithmic (\(V \propto nlog(u)\)) \cite{briscoe_metcalfes_2006}, quadratic (\(V \propto u^2\)) \cite{metcalfe_metcalfes_2013} or other (e.g. \(V \propto 2^u\)).

Network effects have found application in numerous industries and business models ranging from telecommunications \cite{birke_network_2004, gallaugher_understanding_2002}, to web servers, PC software \cite{gandal_competing_1995}, airline reservation systems, ATMs \cite{economides_competition_1992}, and platform systems \cite{church_platform_2005}. Indeed, the intuition and implications of network effects have been so pervasive that they have been invoked in any industry where the consumption or use of a product by consumers makes the product more valuable for others (for a collection of relevant literature see \cite{garcia-swartz_network_2015}. It is no surprise that cryptoassets have also been hypothesized to exhibit network effects. The combination of the inherent network nature, the meteoric rise in popularity (read: userbase), and the substantial price volatility (read: value) has suggested a strong-if elusive-relationship. 

The particular motivation behind the study of network effects in cryptoassets has so far been to discover a valuation formula: if we know the function between userbase and value, then with informed guesses on the network's growth we can predict future prices \cite{peterson_metcalfes_2018, van_vliet_alternative_2018, shanaev_marginal_2019}. But valuation formulas reduce network effects down to a binary distinction represented by a single function. While useful as prediction tools and high-level descriptors of cryptoasset trends, valuation formulas provide little granularity.

Our motivation and goal is, instead, to provide more high-level view of how network effects influence the cryptoasset market as a whole, and particularly what they say about the potential for concentration in the market and about competitive (dis)advantages of one cryptoasset over others. These are the most impactful implications of network effects, and they are desirable for those networks that can exploit them, but undesirable for their competitors or for regulators who have to deal with concentrated markets. We work with numerous cryptoassets so that we can obtain a market-wide overview (limited by how big and representative our sample is), and we study them from their inception until early 2020 which allows us to capture all historically important phases, including the resurgence in 2019, which extant literature has not had a chance to consider. This type of approach allows us to draw insights about the structure and competitive dynamics of the cryptoasset market. It goes back to the early wave of "\textit{Bitcoin maximalism}", which stood for the idea that the optimal number of currencies as alternatives to the mainstream financial system is one, and altcoins will eventually be rendered obsolete as more and more users gravitate toward the biggest, most stable, most widely accepted cryptocurrency, namely Bitcoin. At the time, Bitcoin maximalism was rejected by Vitalik Buterin, the creator of Ethereum, correctly pointing out that the cryptoasset universe is not a homogeneous thing, and that therefore there is no one single "network" around which network effects would form \cite{buterin_bitcoin_2014}. We expand on that thinking.

Looking at network effects to study the competitive dynamics of the cryptoasset market and its potential to concentrate around one or a small number of cryptoassets can provide useful insights for industrial policy. Normally, a showing that cryptoassets exhibit network effects would suggest that early cryptoassets have a first-mover advantage and may lock the market in \cite{economides_economics_1996, katz_network_1985, gandal_can_2016}, even if they are intrinsically inferior to other comparable cryptoassets \cite{farrell_standardization_1985,hagiu_network_2016,briscoe_metcalfes_2006}. While, the market seems to have moved away from that danger, network effects theory also suggests that, assuming homogeneity, once a cryptoasset hits a tipping point, it may fully prevail because new users will always prefer the cryptoasset with the larger userbase (the so called "winner-take-all" markets, which Bitcoin maximalism relied on) \cite{economides_economics_1996, katz_network_1985}. Homogeneity is, of course, a matter of degree, and it is still likely that, if a cryptoasset exhibits stronger network effects than its peers, it can prevail at least within a sub-segment of the market. The flip side of network effects can also be observed, whereby the loss of a user results in a supra-proportionate loss of value (i.e. more value than the user intrinsically contributed individually), which incites further losses and so on. This means that rapid depreciation is more likely in cryptoassets characterized by network effects. The rapid appreciation and depreciation cycles coupled with the winner-take-all characteristic can in turn result in cryptoasset markets that are successively dominated by a new winner in every era (successive contestable monopolies). Then, if this is the natural state of the market, artificially forcing more competition may not be optimal.

These insights are well-applicable in financial markets. For instance, the influential "Cruickshank report", an independent report on banking services in the United Kingdom prepared for the UK Treasury, which has in turn influenced regulatory and legal decisions \cite{noauthor_morgan_2007, noauthor_competition_2011}, warned about the far reaching implications of network effects: "Network effects also have profound implications for competition, efficiency and innovation in markets where they arise. Establishing critical mass is the first hurdle, as the benefits to customers and businesses of a network arise only gradually with increasing use. It is possible to imagine a world in which electronic cash is widely held and used, for example, but much harder to see how to get there. Once a network is well established, it can be extremely difficult to create a new network in direct competition.  ... Where network effects are strong, the number of competing networks is likely to be small and the entry barriers facing new networks will be high" \cite{cruickshank_competition_2000}. As the fintech industry is heating up, network effects have also been cited there as a strong factor in entrenching existing market power of financial services (see e.g. the recent proposed acquisition of Plaid by Visa \cite{cyphers_visa_2020}), and such risks have also been highlighted in the cryptoasset market, with models showing that certain conditions can allow cryptoasset markets to become oligopolies and market players entrench their position in the market \cite{arnosti_bitcoin_2018, cong_tokenomics_2020}.

\section{Prior Literature and Contribution}
A number of papers have investigated aspects of the application of network effects in cryptoasset networks. The focus has been to determine whether the value of cryptoassets (and mainly Bitcoin) complies with network effects, and in particular on whether it follows Metcalfe's law, which is the most popular iteration of network effects and stipulates that the value of a network grows at a rate proportional to the square of the number of users (\(V \propto u^2\)).

The early influential analysis by Peterson \cite{peterson_metcalfes_2018}  remains the point of reference. Peterson developed a valuation model for Bitcoin's price based on Metcalfe's law for the period 2009-2017, using wallets as a proxy for users, Bitcoin prices as the proxy for value, and a Gompertz function to account for growth. He  found that the price of Bitcoin follows Metcalfe's law with R-square of 85 percent. In a revised version of the original paper that extends through 2019, Peterson re-confirms the application of Metcalfe's law to Bitcoin \cite{peterson_bitcoin_2019}. However, he excludes significant periods of time on the grounds of price manipulation, during which the value of the Bitcoin network, as measured by the USD price of Bitcoin, lies well outside of Peterson's model predictions. Van Vliet \cite{van_vliet_alternative_2018} enhanced Peterson's model by incorporating Rogers' diffusion of innovation models to better capture population parameters and growth rates. By doing so, van Vliet raised R-squared to 99 percent. Shanaev et al. \cite{shanaev_marginal_2019} acknowledge the utility of Peterson's and van Vliet's analyses but depart from them in that their model does not rely on historical data for the estimation of the coefficient of proportionality, which raises an endogeneity problem. They still use Metcalfe's law but only as only as one of the building blocks of their model. Civitarese \cite{civitarese_does_2018} rejects the applicability of Metcalfe's law to the value of the Bitcoin network by running a cointegration test between price and an adjusted number of wallets' connections.

Gandal and Halaburda \cite{gandal_can_2016} use a completely different approach to examine the existence of network effects in cryptoasset networks. They define network effects as the reinforcement effects the price of a cryptoasset has on the price of another cryptoasset. With Bitcoin as the base cryptoasset, the idea is that, if network effects are in place, as Bitcoin becomes more popular (price increase), more people will believe that it will win the winner-take-all race against other cryptoassets resulting in further demand and higher prices. Therefore, network effects would manifest themselves as an inverse (negative) correlation between the prices of the sampled cryptoassets. For the period May 2013 - July 2014, their results showed signs of network effects after April 2014.

Our analysis complements and differs from prior literature in several ways. Firstly, we do not focus on a specific network effects formula; we rather look at when, to what degree, in which cryptoassets, and for what proxies of value and userbase network effects are observable (defined as supra-proportional change in value relative to userbase) regardless of which particular curve/function they follow. Secondly, we go beyond Bitcoin to examine six cryptoassets that we have selected as representative of different features and characteristics to better be able to observe potential industry-wide trends. This helps us notice whether one cryptoasset has the potential to dominate the market or multiple cryptoassets benefit from the same network effect forces. Thirdly, we use different parameters as proxies for value and userbase to more fully capture the functionality and usage of cryptoassets in the market. Importantly, we do not rely on the total number of users as a proxy for userbase like extant literature, because many of those addresses are dormant or permanently inaccessible and therefore economically irrelevant. Fourthly, we study the full history of cryptoassets from their inception to today which allows us to observe their different phases, including the price collapse in 2018 and the resurgence in mid-2019, which dramatically change the picture of network effects and which have been missed by previous studies. Lastly, we work with data sets that have been meticulously cleaned to filter out spurious or manipulative activity, which improves the accuracy of our results compared to data-sets that are pulled unfiltered from the network. Our analysis confirms the existence of network effects, but also that they do not have the results usually associated with them on the market.

\section{Methodology and Development}
We study the application of network effects in Bitcoin (BTC), Dogecoin (DOGE), Ethereum (ETH), Litecoin (LTC), XRP and Tezos (XTZ). The selection of these cryptoassets was made on the basis of diversity and feasibility. We aimed to study cryptoassets that exhibited different attributes in terms of age, market capitalization and any special features that make them stand out from other competing cryptoassets in order to build a representative sample of the crypto-economy \cite{irresberger_public_2020}. We also limited the study to cryptoassets for which we could get reliable, standardized time-series data from the cryptoassets' initial release to the time of the study \cite{coin_metrics_coin_2020}. The unreliability of the prices reported by exchanges in the early days of the industry led us to consider Bitcoin from July 2010, Litecoin from March 2013, and XRP from August 2014—the rest from their beginning. Table 1 summarizes the attributes of each chosen cryptoasset. 

\begin{table}[t]
\small
\begin{tabularx}{\columnwidth}{|X|l|l|X|}
\hline
                  & \textbf{Age}  & \textbf{Market cap (2020)} & \textbf{Features}             \\ \hline
\textbf{Bitcoin  (BTC)}  & Old (2009)    & V. Large (\$170B)        & Popularity, first cryptocurrency, UTXO based                           \\
\textbf{Dogecoin (DOGE)} & Old (2013)    & V. Small (\$0.3B)            & "Joke cryptocurrency", early BTC contender , UTXO based\\
\textbf{Ethereum (ETH)} & Medium (2015) & Medium (\$25B)             & Turing complete, programmable, account based            \\
\textbf{Litecoin (LTC)}    & Old (2011) & Small (\$2.6B)            & First major BTC fork, UTXO based                 \\
\textbf{XRP}   & Old (2012)    & Small (\$8B)             & Consensus, fintech-orientated, account based        \\
\textbf{Tezos (XTZ)}    & New (2018)    & Small (\$1.7B)             & Centralized PoS, on chain governance, account based \\ \hline
\end{tabularx}
\vspace{2mm}
\caption{List of studied cryptoassets, chosen to cover different characteristics.}
\label{table:crypto_overview}
\end{table}

We first define network effects. Network effects occur where the value of the network \(V\) grows supra-proportionately to the number of users \(n\) that participate in the network. Reverse network effects occur where the value \(V\) drops supra-proportionately to the number of users \(n\) that leave the network. Unless there is a reason to distinguish between positive and reverse network effects, we collectively refer to them as network effects. Therefore, we define network effects to occur in cryptoassets when a positive value change $\Delta V > 0$ is larger than a positive userbase change $\Delta u > 0$, or when a negative value change $\Delta V < 0$ is smaller than a negative userbase change $\Delta u < 0$. Notice that we do not consider that network effects apply when value and userbase move in different directions, e.g. when the value increases while the userbase decreases, regardless of which increases or decreases more. 

Thus, network effects occur if
\[\Delta V > \Delta u \geq 0\ \lor \Delta V < \Delta u \leq 0\]

In our analysis we define change at time $t$ similar to log returns, i.e.
\begin{align}
    \Delta V &:=\ln \frac{V_{t+1}}{V_t} \\
    \Delta u &:= \ln \frac{u_{t+1}}{u_t}
\end{align}

Then, we identify appropriate proxies to represent value \(V\) and userbase \(u\). To represent \(V\) we use two proxies: (a) token price and (b) transaction value. The two proxies represent different aspects of the value users assign to cryptoassets. In theory, even one proxy applied to one cryptoasset would be enough to demonstrate (or not) network effects (as has, for example, been done in previous literature that relied only on token price), assuming the proxy and cryptoasset are representative. However, because cryptoassets are differentiated resulting in diversified usage patterns, and because the chosen proxies express different ways by which users perceive the value of the network, a multitude of cryptoassets and proxies was used in an effort to better represent the industry.

\textbf{Token Price (PriceUSD): }The first parameter we use is token price, which is the fixed closing price of the asset as of 00:00 UTC the following day (i.e., midnight UTC of the current day) denominated in USD (for a detailed explanation of Coin Metric's methodology on toke price see \cite{coin_metrics_coin_2020}). Token price expresses value in terms of market forces, namely the point at which supply meets demand. It is the value that users as market participants collectively assign to a given cryptoasset by deciding to buy and sell at that price level. We assume that the studied cryptoassets trade under normal market conditions; any acknowledgement of price manipulation that may have occurred at times has been accounted for in the cleaning of data by Coin Metrics \cite{coin_metrics_coin_2020}.

\textbf{Transaction Value (TxTfrValAdjUSD): }The second proxy of choice is transaction value, which expresses the USD value of the sum of native units transferred between distinct addresses per day removing noise and certain artifacts to better reflect the real economically relevant value circulating in the network. The assumption is that as the network becomes more valuable to users, they will use it more frequently and/or to transfer greater value among them. Therefore, transaction value as a proxy sees cryptoassets as means of transaction. We considered and rejected transaction count as an appropriate proxy, because on some networks a large number of recorded transactions are unrelated to value transfer, but rather to the operation of the network, e.g. consensus formation on Tezos \cite{perez_revisiting_2020}. One could retort that even these non-value-carrying transactions reflect engagement with the network and that therefore are an indication of the value of the network to users. Even so, lumping together value-carrying and operational transactions would taint the comparison across cryptoassets, since on some cryptoassets the majority of transactions are operational (e.g. Tezos, see \cite{perez_revisiting_2020}), while on others value-carrying (e.g. Bitcoin). 

Next, to represent \(u\) we select the following proxies: (a) addresses with non-zero balance (b) trailing 6-month active addresses and . Different ways to represent userbase more fully captures the relationship between value and userbase. We considered and rejected counting userbase based on total number of addresses (like all previous literature), because of the large number of inactive addresses. Contrary to other industries where network effects have been studied and where inactive users are eventually purged from the network (e.g. mobile phone subscriptions, social networks), so that total user count may still be a good approximation of the economically meaningful userbase, this is not the case with cryptoassets. Instead we opted for two variants of addresses with non-zero balance, as defined below. 

\textbf{Addresses with Non-Zero Balance (AdrBalCnt):} This proxy represents the sum count of unique addresses holding any amount of native units as of the end of that day. Only native units are considered (e.g., a 0 ETH balance address with ERC-20 tokens would not be considered). The utility of this proxy lies in that it excludes all non-economically active addresses, the assumption being that addresses with zero balance are dormant (similar to bank accounts with zero balance). This choice responds to criticism that has been raised with regard to extant literature that tended to use all addresses or wallets as a proxy for users. Despite our choice of improved metric, it still remains a fact that there is no one-to-one mapping between addresses and actual users, which is a common problem to any network or service, e.g. the same person may have multiple bank accounts. While there are methods to de-cluster actual users from wallets and addresses, these are not sufficiently precise and are unavailable or inapplicable across cryptoassets \cite{Harrigan_cluster_2020}. We also acknowledge that on networks with lower transaction fees it is easier to generate and/or maintain addresses with balance, and to counter that we could raise the amount of native units the counted addresses should have, but this would introduce a subjectivity question without even fully eradicating the initial problem of spurious addresses.

\textbf{Trailing 6-Month Active Addresses (6MAdrActCnt):} This proxy counts all unique addresses that have been active at least once over the trailing 6-month period from the time of measurement. Repeat activity is not double-counted. Traditionally, most userbase measurements are taken in time frames that range from one month to one year. Given that cryptoassets are of relatively young age, which may suggest that their userbase is expected to interact with them less frequently, and that part of their utility involves simply owning them, which does not generate any activity, we decided that a 6-month time frame sufficiently captures active userbase.

Before we derive network effects, we first calculate the Pearson correlation between value \(V\) and users \(u\) which is informative in terms of their overall relationship. Next, we obtain relevant measurements of network effects. We rely predominantly on the PriceUSD-AdrBalCnt pair of proxies for value and userbase, but additional measurements are in the Appendix. To see how prevalent network effects are in the studied cryptoassets we calculate the ratio of total days to the days where network effects were observed (separately for positive and reverse) for each cryptoasset. To see how strong network effects are we calculate the ratio of total days to the sum of the network effects observations over the days they occurred for each cryptoasset (separately for positive and reverse). To see how strong network effects are in cryptoassets relative to each other we reduce to a 100 day period. The results are presented in Part 5 and the analysis of the results in Part 6. 

\begin{table}[]
\begin{tabular}{@{}ll@{}}
\toprule
Metric abbr & Metric meaning \\ \midrule
PriceUSD    & Token price \\
TxTfrValAdjUSD    & Transaction value \\
AdrBalCnt    & Addresses with non-zero balance \\
6MAdrActCnt    & Trailing 6-month active addresses \\ \midrule  
NFX    & Network effects \\\bottomrule
\end{tabular}
\caption{Legend of metrics in use.}
\end{table}

\section{Results}

We are looking for network effects in the relationship between value \(V\) and users \(u\) of various cryptoassets as represented by the proxies defined previously. Four pairs (2x2 proxies) are possible: 

\begin{itemize}
\item {\verb|Token Price - Addresses with Non-Zero Balance|}: This pair demonstrates network effects expressed as the change of monetary value of a cryptoasset relative to the users that hold any amount of that cryptoasset. By counting only accounts with non-zero balance, we filter out economically dormant users.
\item {\verb|Token Price - Trailing 6-month Active Addresses|}: This pair demonstrates network effects expressed as the change of monetary value of a cryptoasset relative to the users that have been active at least once in the trailing 6-month period on that cryptoasset's network. Counting all active users over a recent time segment (usually 1, 6 or 12 months) is a common measurement of network or platform userbase and less conservative than daily active users.
\item {\verb|Transaction Value - Addresses with Non-Zero Balance|}: This pair demonstrates network effects expressed as the change of transaction value of a cryptoasset relative to the users that hold any amount of that cryptoasset. 
\item {\verb|Transaction Value - Trailing 6-month Active Addresses|}:  This pair demonstrates network effects expressed as the change of transaction value of a cryptoasset relative to the users that have been active at least once in the trailing 6-month period on that cryptoasset's network.
\end{itemize}

Before we derive network effects, we calculate, based on the above pairs, the Pearson correlation between value \(V\) and users \(u\) which tells us whether, as a general matter, cryptoasset value and userbase are moving in the same direction. This already provides an indication of whether cryptoassets become more valuable as their adoption increases.

\begin{table}[]
    \centering
    \begin{tabular}{llrr}
\toprule
& & \multicolumn{2}{c}{User proxies} \\
\cline{3-4}
Cryptoasset &      Value proxy &  AdrBalCnt &  6MAdrActCnt \\
\midrule
     BTC &        PriceUSD &   0.878760 &     0.800890 \\
     BTC &  TxTfrValAdjUSD &   0.771601 &     0.734617 \\
    DOGE &        PriceUSD &   0.532856 &     0.255025 \\
    DOGE &  TxTfrValAdjUSD &   0.258791 &     0.141790 \\
     ETH &        PriceUSD &   0.256837 &     0.475199 \\
     ETH &  TxTfrValAdjUSD &   0.048093 &     0.214427 \\
     LTC &        PriceUSD &   0.646814 &     0.844012 \\
     LTC &  TxTfrValAdjUSD &   0.258648 &     0.431706 \\
     XRP &        PriceUSD &   0.551157 &     0.803027 \\
     XRP &  TxTfrValAdjUSD &   0.189622 &     0.278429 \\
     XTZ &        PriceUSD &  -0.477943 &    -0.681394 \\
     XTZ &  TxTfrValAdjUSD &  -0.169407 &    -0.240346 \\
\bottomrule
\end{tabular}
    \caption{Pearson correlation between value and user proxies}
    \label{tab:pearson_table}
\end{table}

It is evident that only BTC shows a strong correlation between value and userbase, at least when userbase is measured by our main proxy of total addresses with non-zero balance (AdrBalCnt), with LTC showing the next highest correlation, which is, however, average and only holds when value is measured as value in fiat currency (PriceUSD). Correlations when userbase is measured as addresses that have been active in the trailing 6-month period (6MAdrActCnt) tend to be higher although still not consistently so. Higher correlation using 6MAdrActCnt might be explained on the grounds that user activity picks up during phases of large price movements. Overall, the mediocre and inconsistent correlations between value and userbase provide a first indication that a blanket conclusion that the cryptoasset market is characterized or not by network effects is unwarranted.

Next, we obtain relevant measurements based on the PriceUSD-AdrBalCnt pair of proxies for value and userbase as presented in Table ~\ref{tab:FX_measurements} (additional measurements for other pairs are in the Appendix). As explained in the methodology, we believe these are the most appropriate proxies. Column 5 of Table ~\ref{tab:FX_measurements} shows prevalence of network effects for each cryptoasset as calculated by the ratio of total days to the days where network effects were observed (separately for positive and reverse). Column 6 of Table ~\ref{tab:FX_measurements} shows strength of network effects as calculated by the ratio of total days to the sum of the network effects observations over the days they occurred for each cryptoasset (separately for positive and reverse). Column 7 of Table ~\ref{tab:FX_measurements} shows relative strength of network effects across cryptoassets by reducing  to a 100 day period. This allows us to compare how strong network effects are across cryptoassets regardless of how prevalent they are across them.

\begin{table*}
\begin{tabular}{lr>{\raggedleft\arraybackslash}p{2cm}>{\raggedleft\arraybackslash}p{2cm}>{\raggedleft\arraybackslash}p{4cm}>{\raggedleft\arraybackslash}p{4cm}}
\toprule \\
\centering Crypto & \centering Total days & \centering Days of NFX (pos-reverse) & \centering Sum (strength) of NFX (pos-reverse) & \centering Ratio of total days/NFX days (pos-reverse) &  Relative strength of NFX (pos-reverse) \\ \midrule
Bitcoin    & 3461       & 1434                         & 47.1      & 0.400          & 3.28                        \\
      &            & 243                          & 9.2       & 0.070         & 3.78                        \\
Doge   & 2175       & 695                          & 30.7      & 0.310         & 4.40                         \\
      &            & 295                          & 11.1      & 0.130         & 3.70                         \\
Ethereum    & 1614       & 707                          & 33.5      & 0.430         & 4.70                         \\
      &            & 12                           & 0.2       & 0.007        & 1.66                        \\
Litecoin    & 2473       & 722                          & 34.5      & 0.290         & 4.77                        \\
      &            & 354                          & 11.2      & 0.140         & 3.16                        \\
XRP    & 1973       & 901                          & 41        & 0.450         & 4.55                        \\
	 &            & -                          & -      &-         & -                        \\
Tezos    & 558        & 244                          & 10.8      & 0.430         & 4.40                         \\ 
	&            & -                          & -      & -         &-                        \\
\bottomrule
 & & & & & \\ 
\end{tabular}
\caption{Network effects measurements based on the Token price - Addresses with non zero balance proxy pair}
\label{tab:FX_measurements}
\end{table*}

\begin{figure*}
    \begin{subfigure}[b]{0.46\textwidth}
        \centering
        \includegraphics[width=\textwidth]{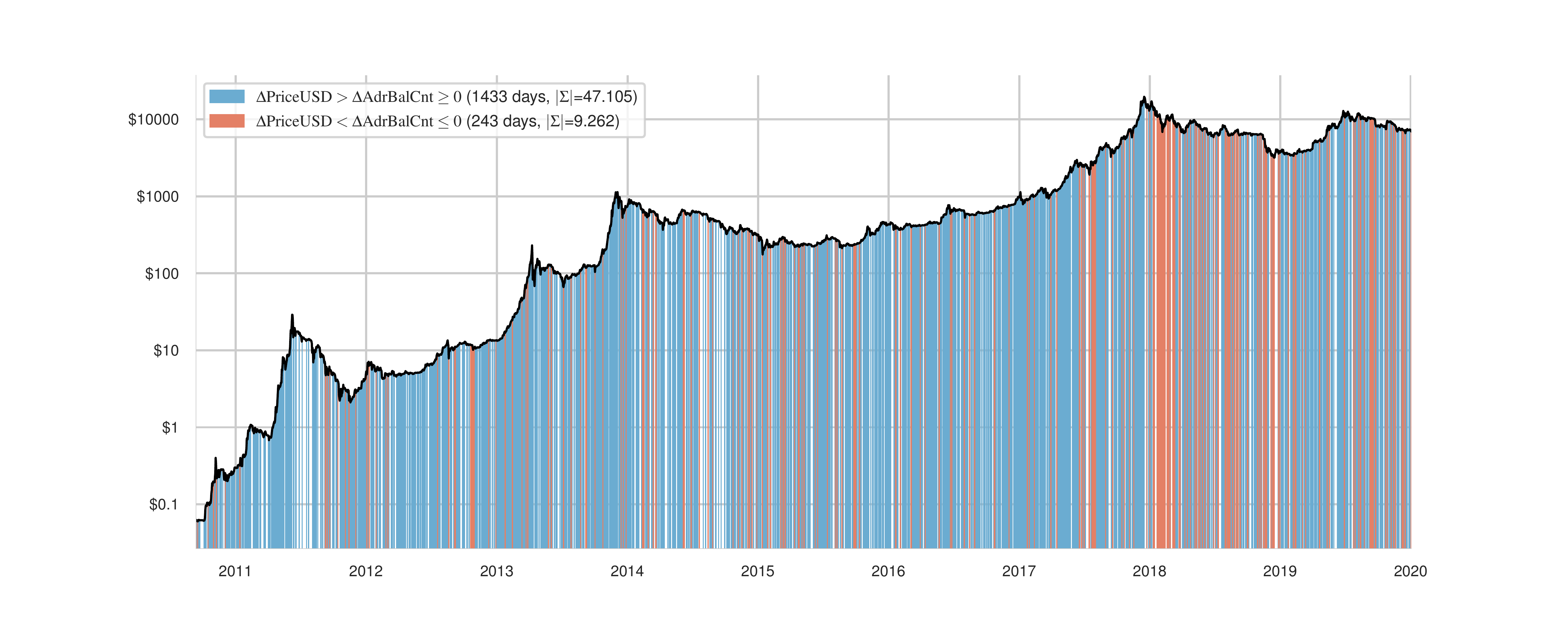}
        \caption{BTC}
        \label{fig:btc-stem}
    \end{subfigure}
    \begin{subfigure}[b]{0.46\textwidth}
        \centering
        \includegraphics[width=\textwidth]{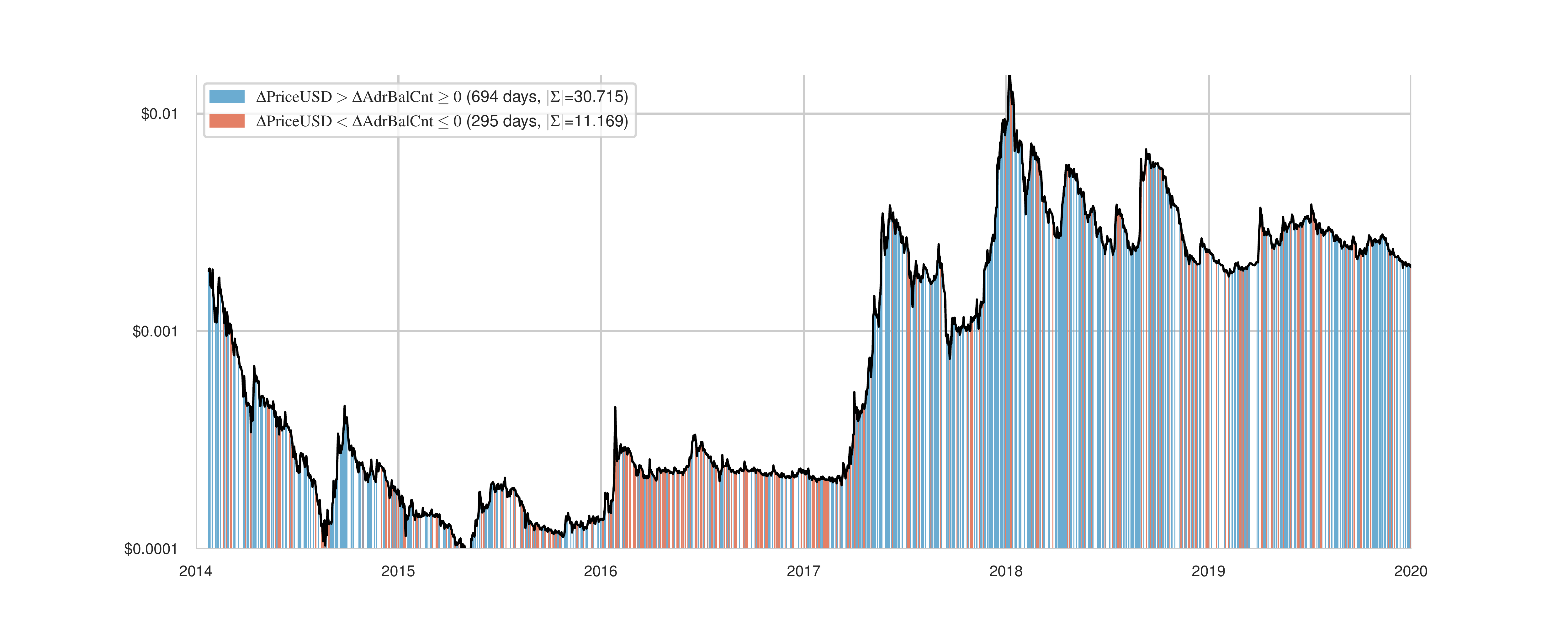}
        \caption{DOGE}
        \label{fig:doge-stem}
    \end{subfigure}
    \begin{subfigure}[b]{0.46\textwidth}
        \centering
        \includegraphics[width=\textwidth]{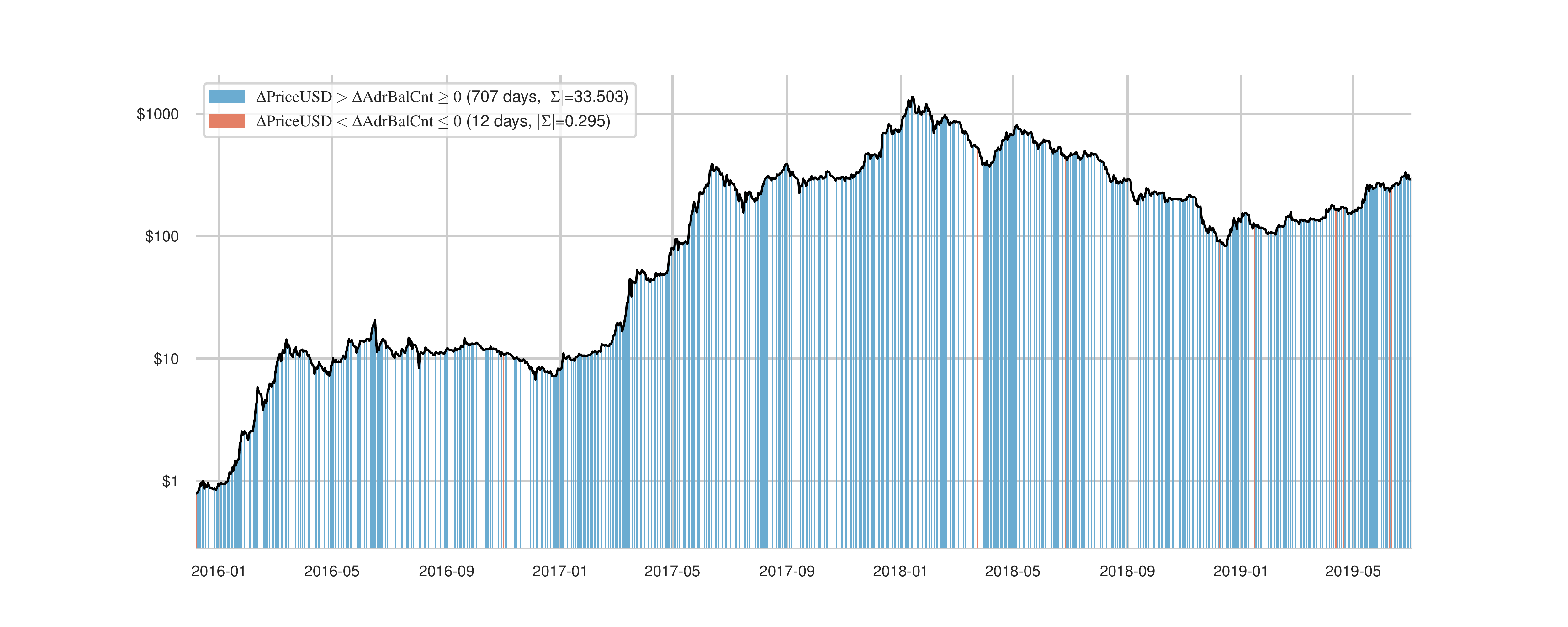}
        \caption{ETH}
         \label{fig:eth-stem}
    \end{subfigure}
        \begin{subfigure}[b]{0.46\textwidth}
        \centering
        \includegraphics[width=\textwidth]{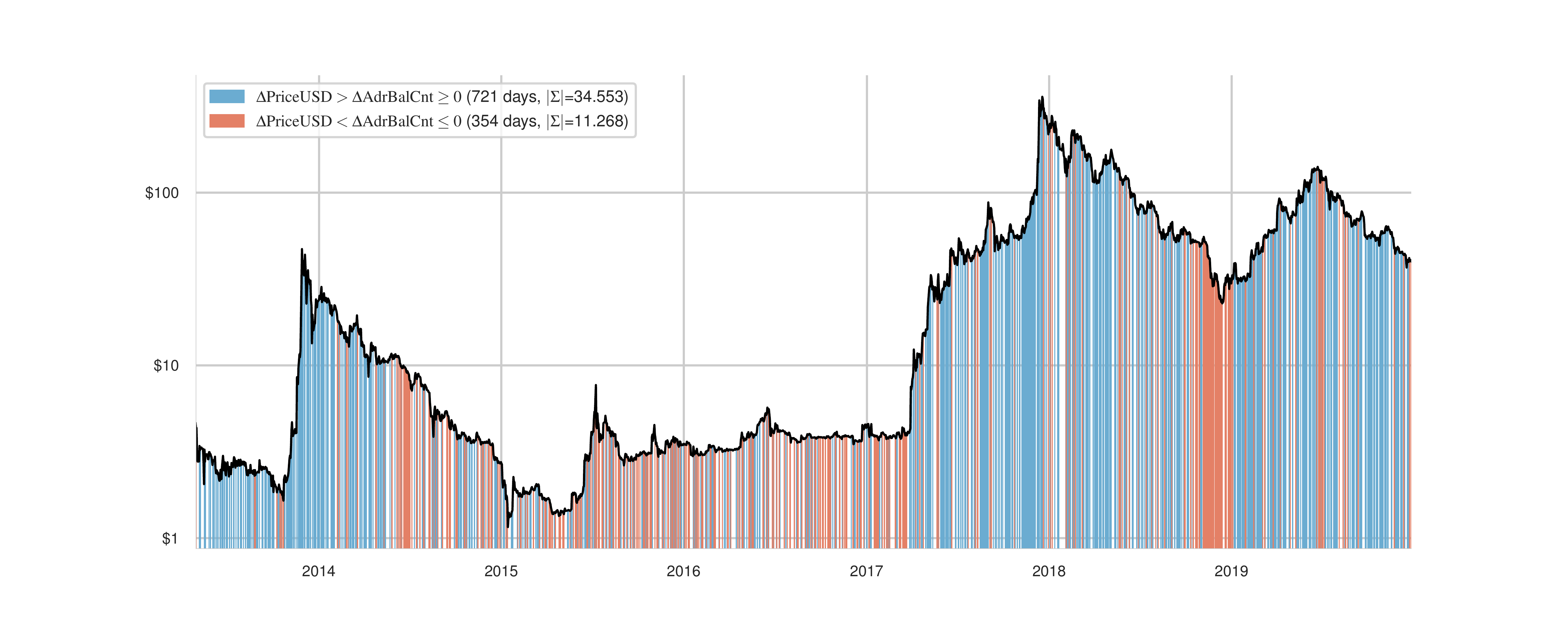}
        \caption{LTC}
         \label{fig:ltc-stem}
    \end{subfigure}
        \begin{subfigure}[b]{0.46\textwidth}
        \centering
        \includegraphics[width=\textwidth]{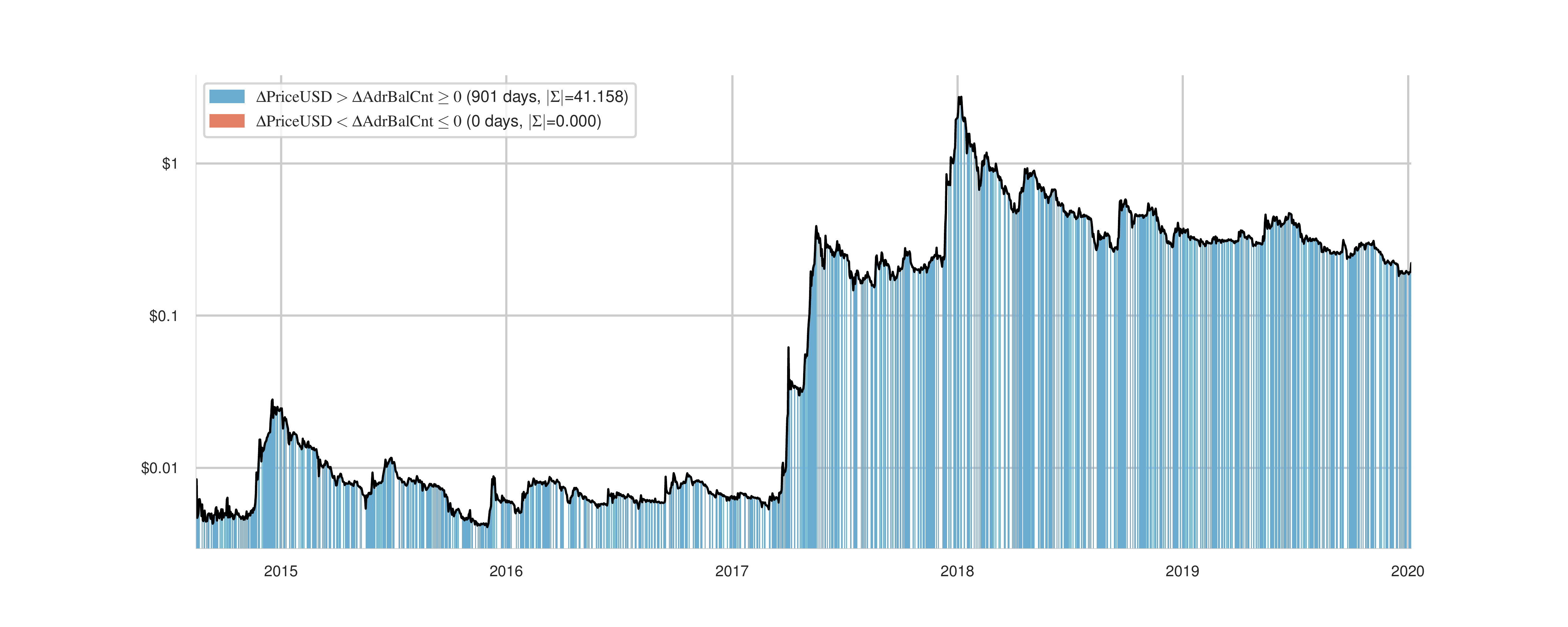}
        \caption{XRP}
         \label{fig:xrp-stem}
    \end{subfigure}
        \begin{subfigure}[b]{0.46\textwidth}
        \centering
        \includegraphics[width=\textwidth]{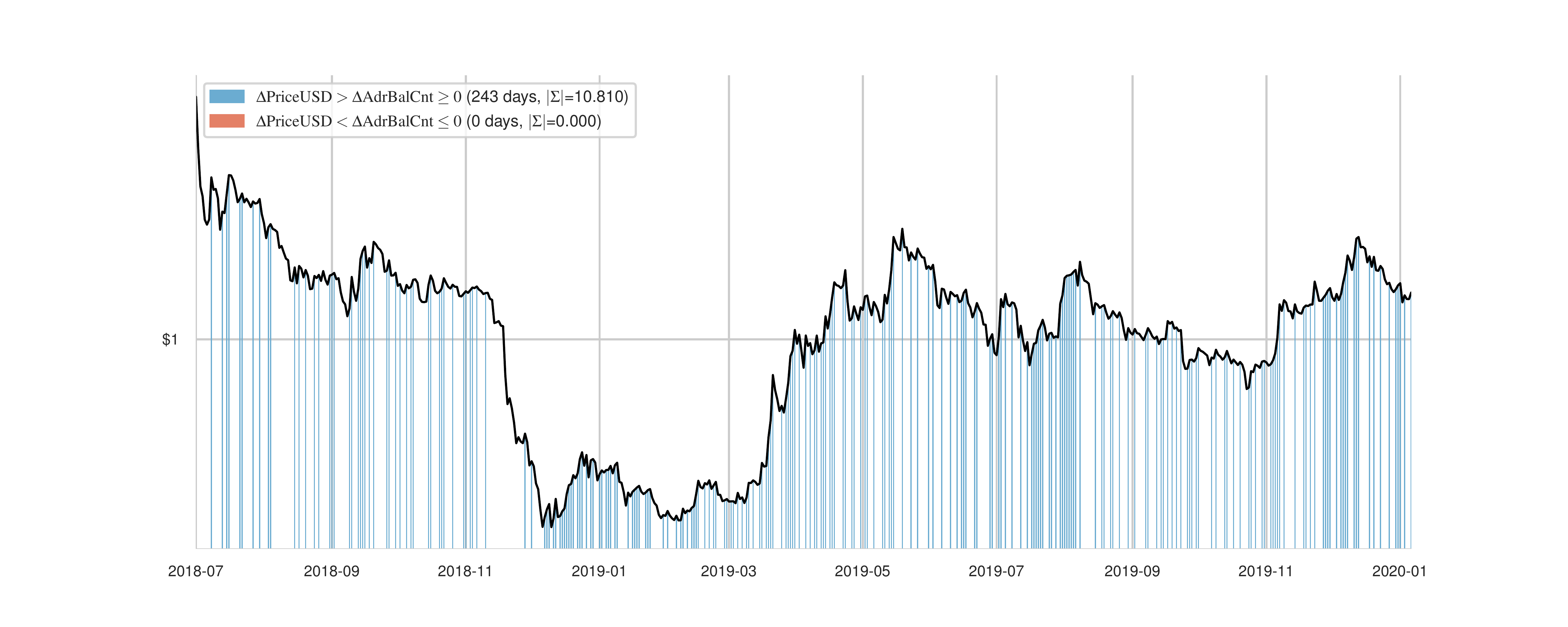}
        \caption{XTZ}
         \label{fig:xtz-stem}
    \end{subfigure}
    \caption{Network effect observations and distribution (blue: positive NFX, red: reverse NFX, white: no NFX); userbase measured by total addresses with non-zero balance, value measured by USD token price.}
    \label{fig:stemplots}
\end{figure*}

\section{Analysis}

Our results are useful in reaching a number of conclusions on how network effects inform the structure and evolution of the cryptoasset market. 

\begin{figure*}
    \begin{subfigure}[b]{0.46\textwidth}
        \centering
        \includegraphics[width=\textwidth]{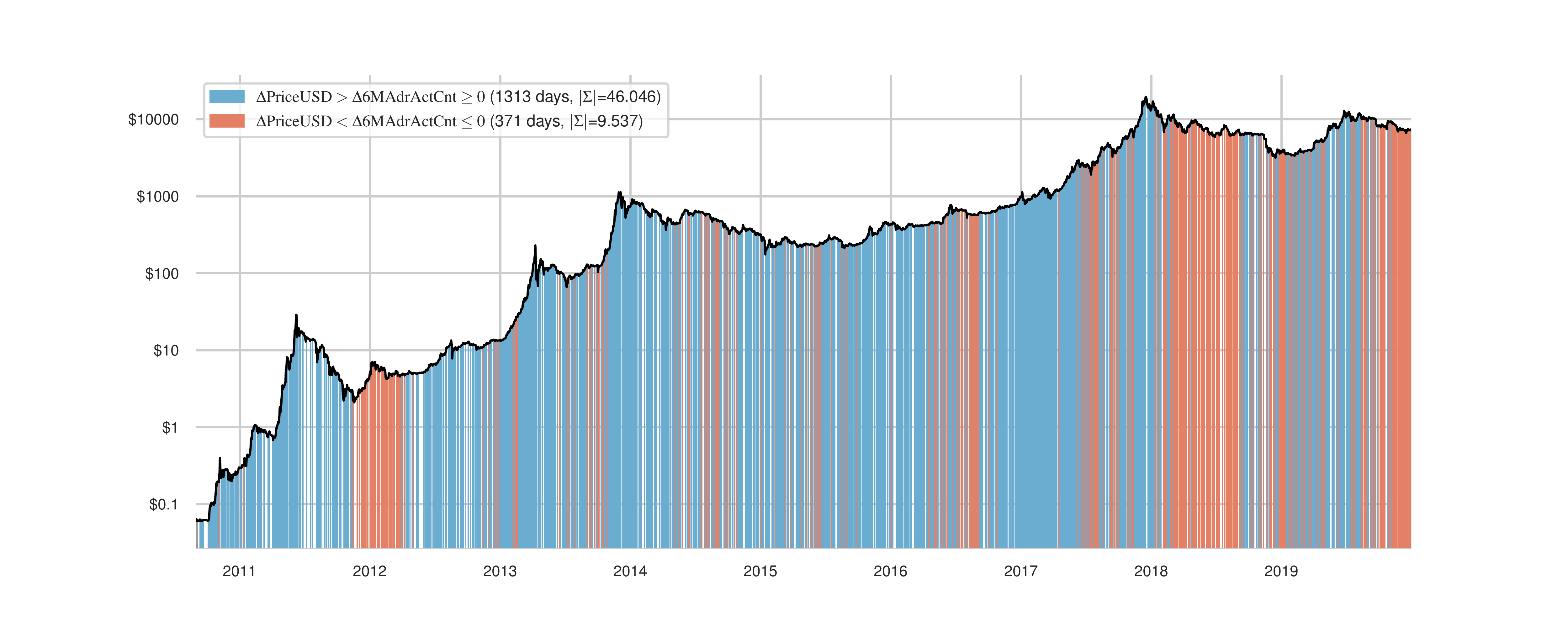}
        \caption{BTC}
        \label{fig:btc-stem-}
    \end{subfigure}
    \begin{subfigure}[b]{0.46\textwidth}
        \centering
        \includegraphics[width=\textwidth]{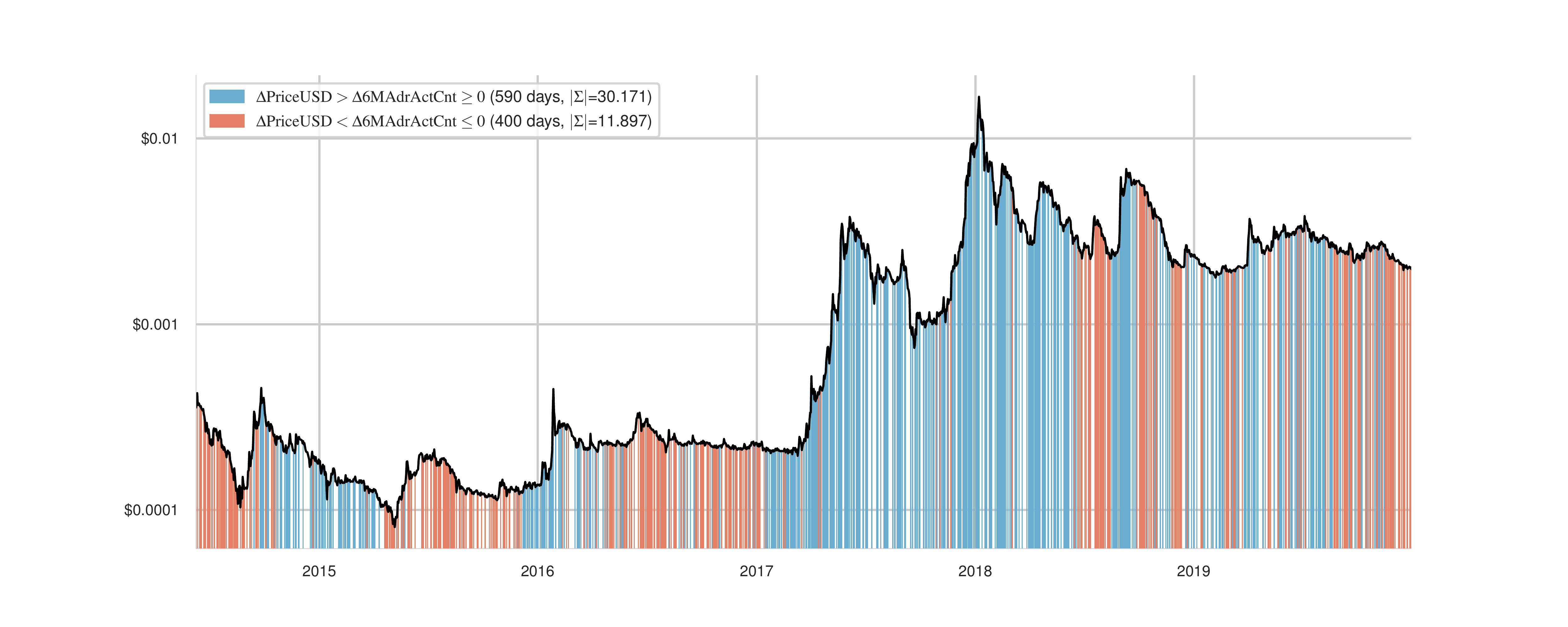}
        \caption{DOGE}
        \label{fig:doge-stem-}
    \end{subfigure}
    \begin{subfigure}[b]{0.46\textwidth}
        \centering
        \includegraphics[width=\textwidth]{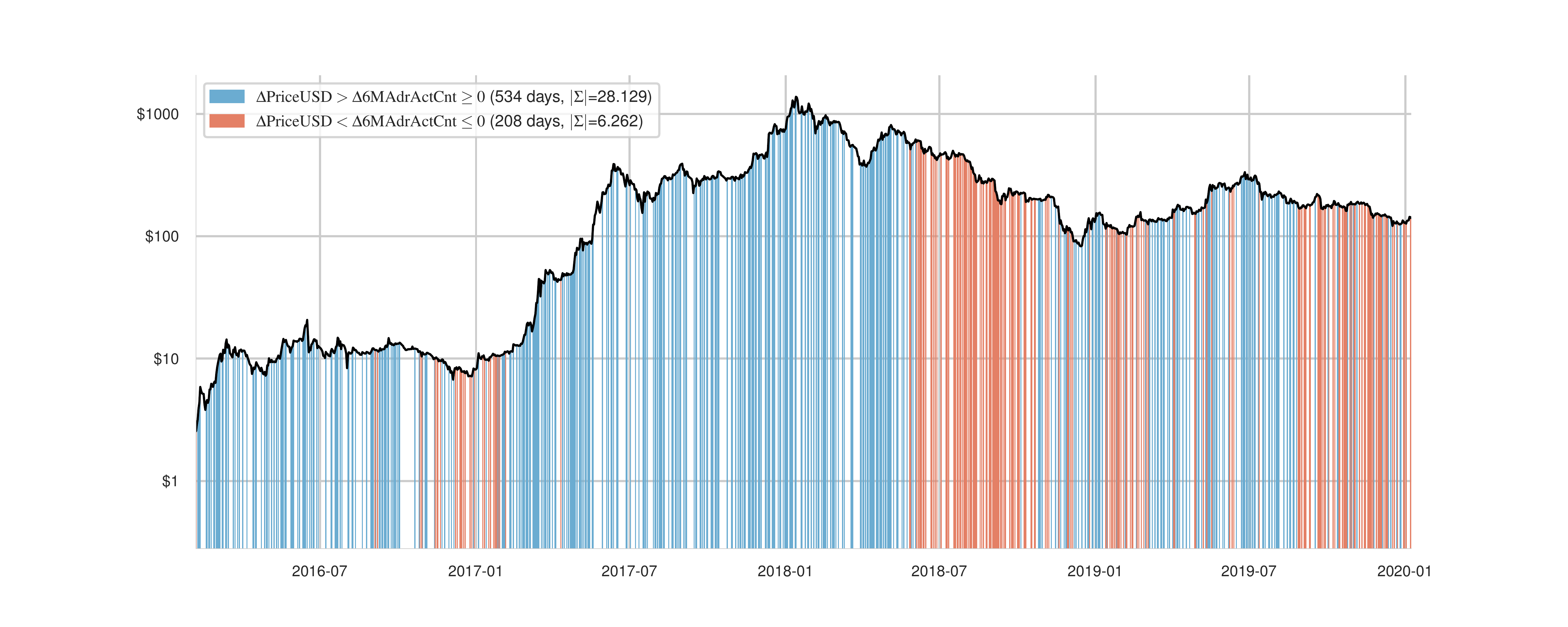}
        \caption{ETH}
         \label{fig:eth-stem-}
    \end{subfigure}
        \begin{subfigure}[b]{0.46\textwidth}
        \centering
        \includegraphics[width=\textwidth]{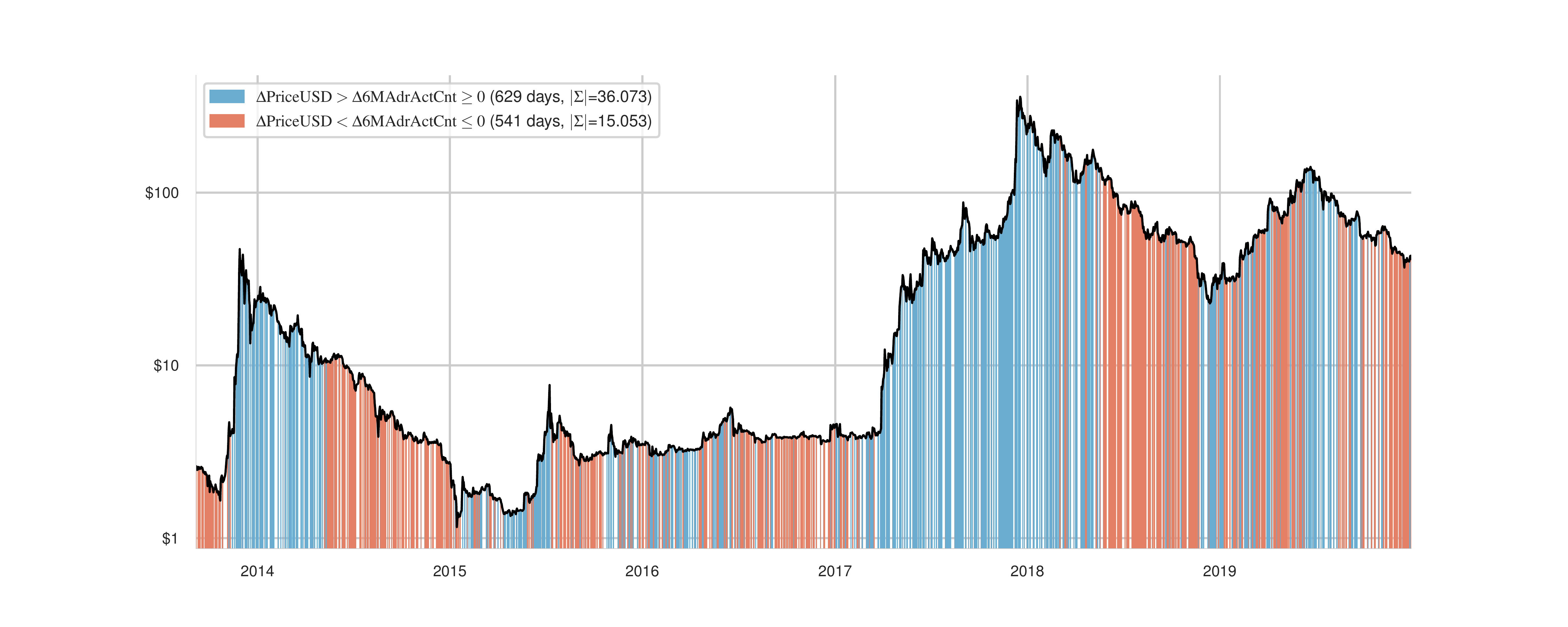}
        \caption{LTC}
         \label{fig:ltc-stem-}
    \end{subfigure}
        \begin{subfigure}[b]{0.46\textwidth}
        \centering
        \includegraphics[width=\textwidth]{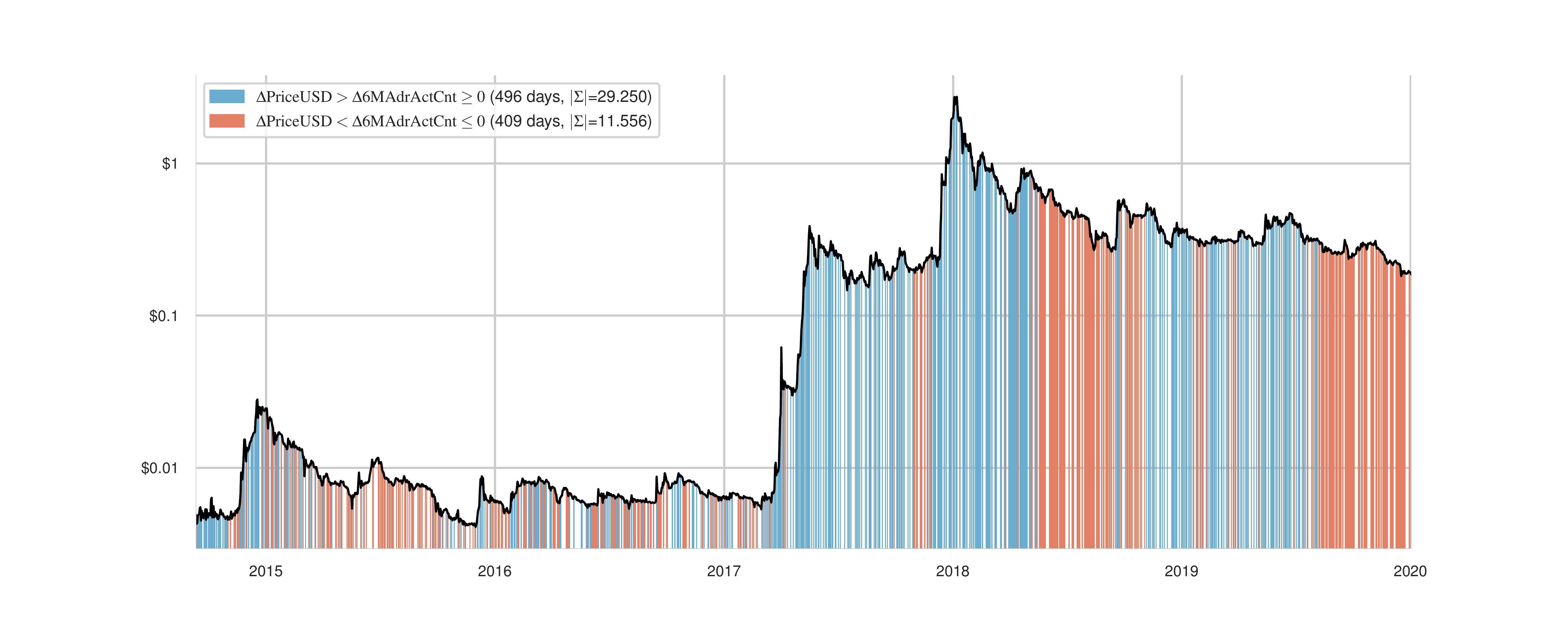}
        \caption{XRP}
         \label{fig:xrp-stem-}
    \end{subfigure}
        \begin{subfigure}[b]{0.46\textwidth}
        \centering
        \includegraphics[width=\textwidth]{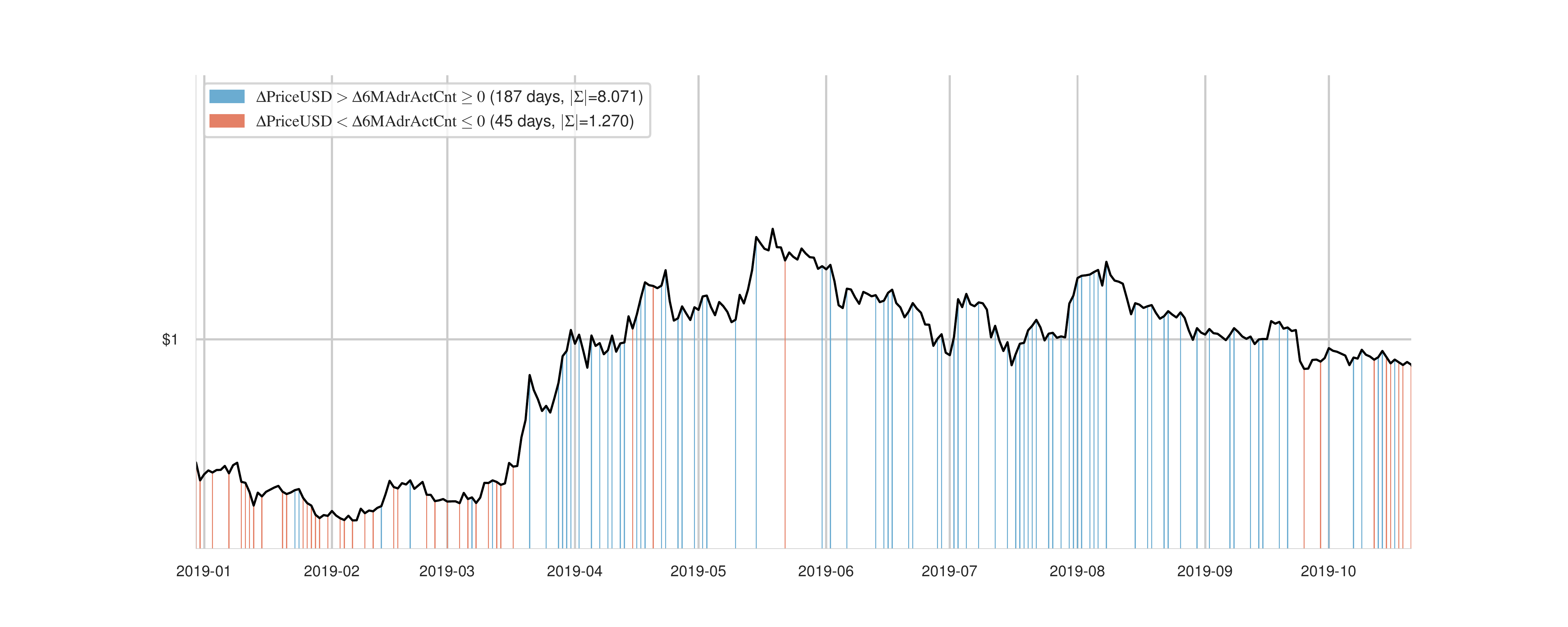}
        \caption{XTZ}
         \label{fig:xtz-stem-}
    \end{subfigure}
    \caption{Network effect observations and distribution (blue: positive NFX, red: reverse NFX, white: no NFX); userbase measured by trailing 6 month addresses, value measured by USD token price.}
    \label{fig:stemplots_6m}
\end{figure*}

\paragraph{(1) Network effects do not provide precise valuation predictions:} 
The most common application of network effects theory has been to draw insights into future cryptoasset pricing based on the evolution of their userbase. Our results indicate that network effect observations in cryptoassets are frequent but inconsistent and therefore they cannot be relied on, generally, as a valuation tool as previous literature suggests (Figures ~\ref{fig:stemplots} and ~\ref{fig:stemplots_6m}). They are most frequent in XRP (45 percent of time in the pair Token Price-Addresses with Non-Zero Balance) and least frequent in LTC (29 percent of time in the same pair). While they appear more consistent in ETH and XRP,  their results can be somewhat misleading at first glance: ETH's and XRP's userbase (AdrBalCnt) was constantly increasing and so any supra-proportionate increase in price registered as a (positive) network effect observation (blue lines in (c) and (e) in Figure ~\ref{fig:stemplots}). However, the positive network effect observations are frequently punctuated by days/periods of no network effect observations during which the price either does not rise supra-proportionately to userbase or drops. In cryptoassets such as BTC and LTC, where userbase fluctuates, it is easier to notice the changes in network effects trends (blue and red lines in (a) and (d) in Figures ~\ref{fig:stemplots} and ~\ref{fig:stemplots_6m}), even through network effect frequency is comparable to ETH and XRP. Therefore, it is hard to conclude that in any cryptoasset network effects exhibit constant patterns that, if extended into the future, can hold predictive value. This does not mean that we do not acknowledge the exponential long-term price increase of some cryptoassets (Figure ~\ref{fig:logdev}), but we note that this is not linked consistently to their userbase growth, which is what network effects theory suggests. One explanation of why our results do not support the conclusions of previous studies can relate to the different time frames. Most previous studies' datasets end around the valuation peak of January 2018, missing the precipitous fall in 2018 and the subsequent rise in 2019, which upend the relatively smoother network effect curves of valuations up until the end of 2017. Another explanation relates to methodology. For example, Peterson's revised study, which covers up to 2020 and confirms the finding of the paper's previous popular version that Bitcoin's valuation follows Metcalfe's law, excludes certain sizeable time periods, which, if accounted for, show a poor(er) fit \cite{peterson_bitcoin_2019}. A third explanation relates to the proxies used. Some previous studies rely on wallets (total addresses) as the proxy for userbase, which is a more crude measurement than our preferred addresses with non-zero balance, as the latter show only economically active users and are therefore a better approximation of relevant userbase.

\paragraph{(2) Reverse network effects are also noticeable meaning that cryptoassets are vulnerable to rapid decline, not just conducive to rapid growth: }
While network effects have mostly been used to describe growth patterns, they are equally applicable in describing decline. Reverse network effects reflect situations where a decrease in users is linked to a larger decrease in value. Such observations are important, because they show that each user loss incurs a greater loss of value and therefore expose the potential for a rapid decline of the network once user exodus begins. Reverse network effects therefore highlight the precariousness of success (as measured by proxies of value). Most cryptoassets exhibited at least one prolonged period where reverse network effects were dominant, during which phases their value contracted disproportionately to the contraction of their userbase ending up less valuable than their userbase size would otherwise suggest or mandate during that period. This is noticeable both when userbase is measured by addresses with non zero balance, but it is even more pronounced when userbase is measured as trailing 6-month active addresses (Figure ~\ref{fig:stemplots_6m}).
This makes sense since the users active in the trailing 6-month period are more likely to be responsive to price fluctuations compared to users who simply hold some balance on their account. From Figure ~\ref{fig:stemplots_6m} it is also evident that user disengagement is almost consistently observed after every price crash (as manifested through the reverse network effects that begin 6 months after many of the crashes), and the fact that price continues to decrease supra-proportionately to userbase, as measured by active users in the trailing 6-month period, 6 months after the crash, may be indicative of the lasting effects user exodus has on the value of cryptoasset networks. Generally, however, while reverse network effects serve as a cautionary note that rapid decline of value can be triggered by user exit, they are weaker in magnitude than positive network effects (Table ~\ref{tab:FX_measurements}). So, overall, positive network effects (albeit inconsistent) still seem to characterize cryptoasset networks.

\paragraph{(3) Cryptoassets do not seem to be a winner-take-all market: }
A common corollary of network effects is that they eventually cause the market to gravitate toward oligopolistic structure, since, everything else equal, users prefer to join the network where the value from their joining will be maximized. This causes a "rich-get-richer" effect where the most valuable network continues to become even more valuable as users prefer to join that over others. Such markets tend to become oligopolistic, with the usual downsides of such industry structure (higher prices, reduced output, entry barriers; lower variety and innovation), and can therefore be a cause for concern. For this to be more likely to happen the various networks (=cryptoassets) must be undifferentiated and switching among and multi-homing across networks must be rare or costly \cite{schmalensee_jeffrey_2011}. These features do not seem to characterize the cryptoasset market, which accordingly appears less susceptible to a winner-take-all trend, at least on account of network effects. Indeed, of the thousands of available cryptoassets many serve different purposes, and users can own multiple cryptoassets at the same time and enter and exit their networks without friction. As evidenced by our results, the fact that the various cryptoassets we studied exhibit network effects of comparable relative strength (Column 7 in Table ~\ref{tab:FX_measurements}), and that they retain their userbase and valuation cycles (Figure ~\ref{fig:logdev}) seems to suggest that the underlying market features, including network effects, do not lead it toward an oligopolistic structure.  

\paragraph{(4) Network effects strength across cryptoassets is comparable and therefore network effects do not accord a single cryptoasset a strong comparative advantage over its peers, undermining fears of concentration: } 
Besides frequency and duration, i.e. what period of a cryptoasset's lifetime is dominated by network effects, another useful parameter of network effect observations in cryptoassets is their strength, i.e. the magnitude of the impact of a userbase change to value change \cite{shankar_network_2003}. Strong network effects can be indicative of higher homogeneity or cohesion within the network, where the addition of each new user (e.g. investor) affects existing users of that closely-knit network more than if it was a different looser network. In turn, this is reflected in the value of the network, or they may be indicative of stronger reputational effects, where the addition of each new user signals major changes for the network, which are then reflected in its value. Our results show that the comparative strength of network effects across the studied cryptoassets is similar (Table ~\ref{tab:FX_measurements}). This leads us to believe that no single cryptoasset benefits from network effects significantly more than its peers and therefore that no cryptoasset enjoys an overwhelming competitive advantage over its peers on account of network effects. A necessary corollary observation is that network effects accrue at similar levels to the studied cryptoassets, which means that network effects as a phenomenon, characterizes the cryptoasset industry as a whole (at least based on our sample), not just Bitcoin, which has been the main subject of many of extant studies in the area. This is not a surprising finding, but it is worth highlighting that it lends support to the previous point that the structure of the cryptoasset market does not seem to be such where network effects lead it to concentration around a small number of cryptoassets or that it helps cryptoassets overtake their peers on account of network effects. This is most likely because cryptoassets are differentiated and multi-homing and switching are pervasive.

\begin{figure*}
    \begin{subfigure}[b]{0.46\textwidth}
        \centering
        \includegraphics[width=\textwidth]{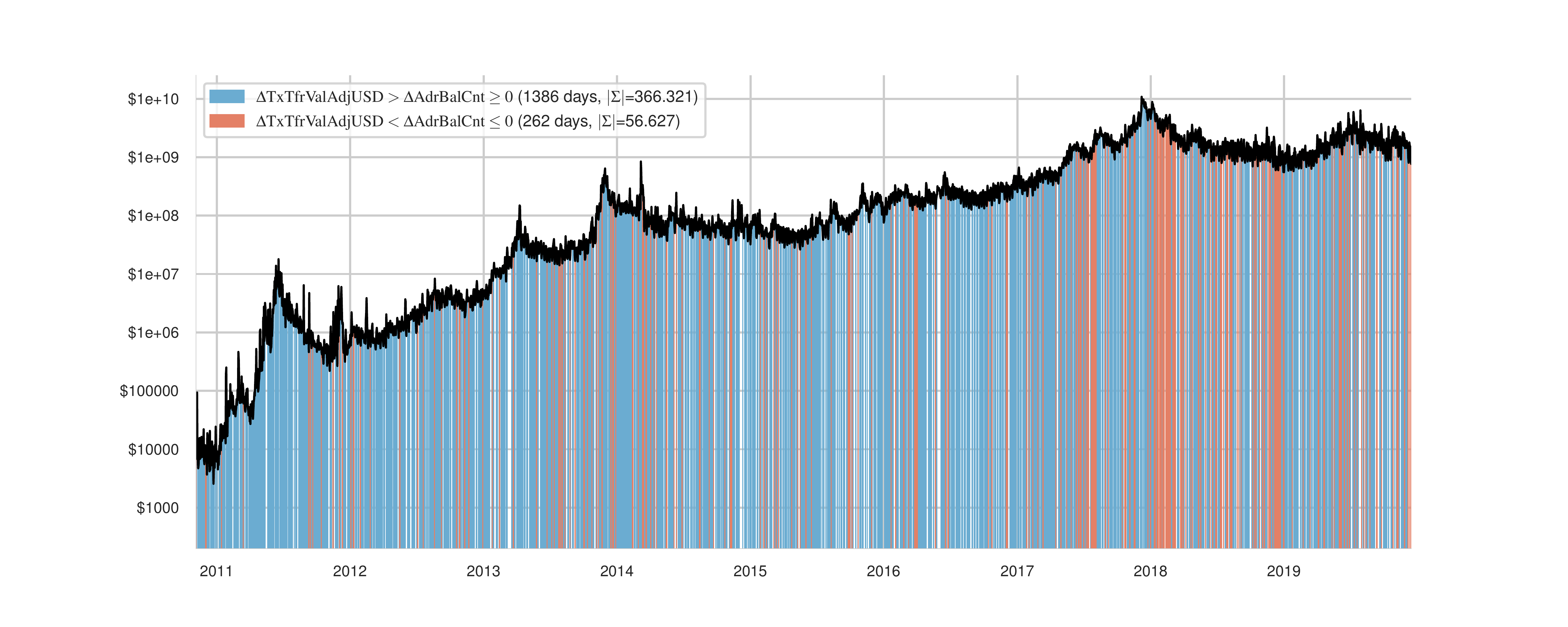}
        \caption{BTC}
        \label{fig:btc-stem_tx}
    \end{subfigure}
    \begin{subfigure}[b]{0.46\textwidth}
        \centering
        \includegraphics[width=\textwidth]{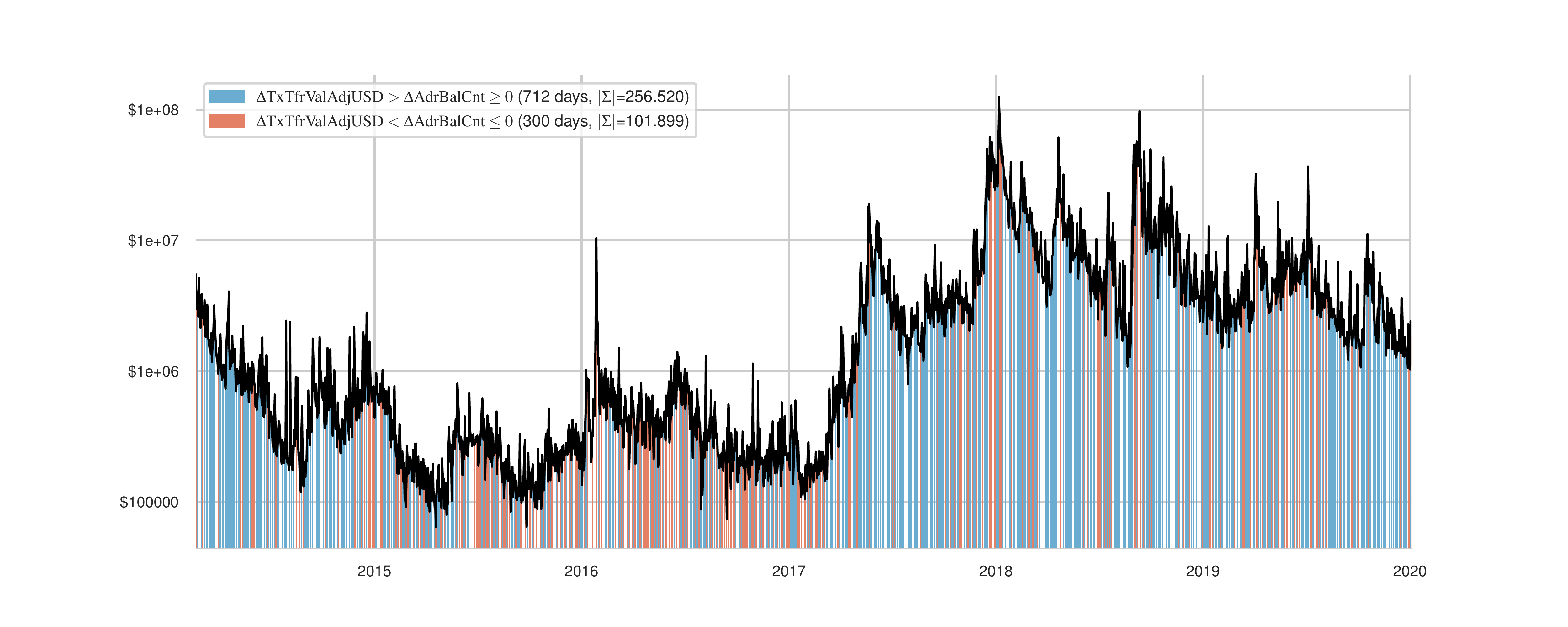}
        \caption{DOGE}
        \label{fig:doge-stem_tx}
    \end{subfigure}
    \begin{subfigure}[b]{0.46\textwidth}
        \centering
        \includegraphics[width=\textwidth]{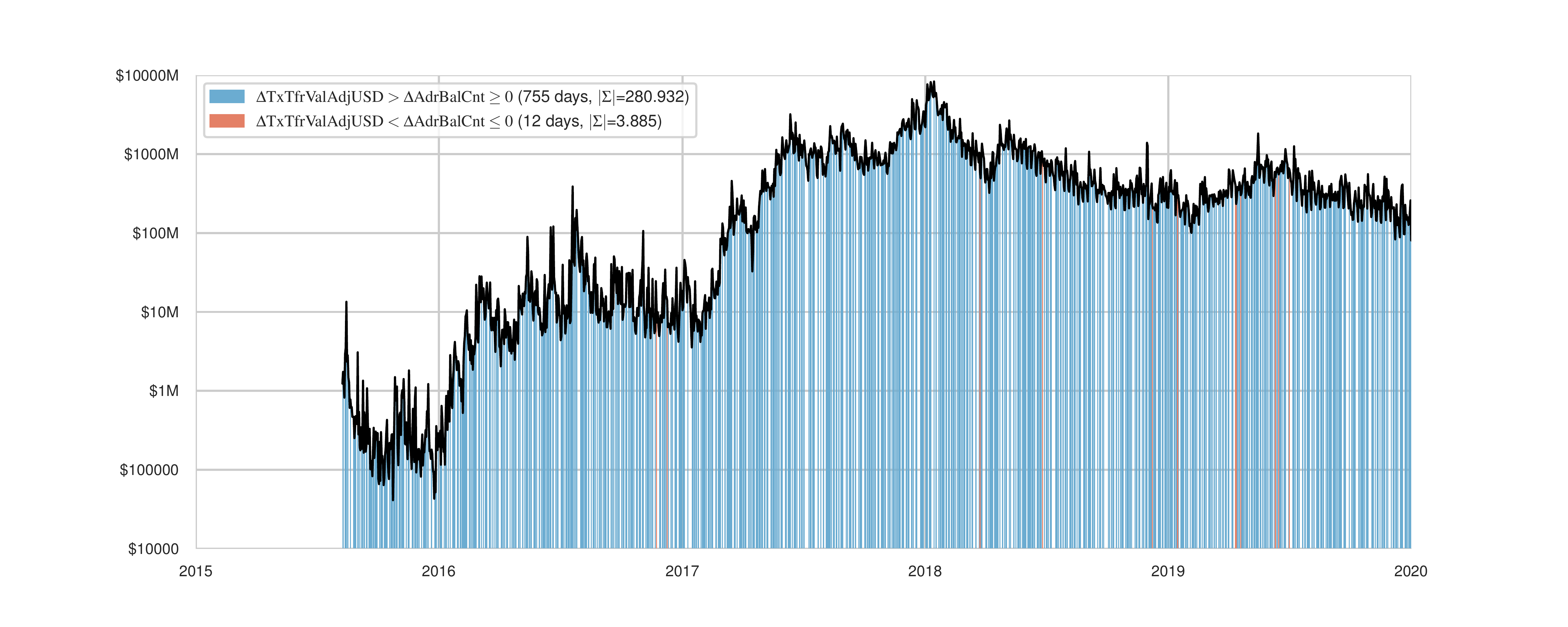}
        \caption{ETH}
         \label{fig:eth-stem_tx}
    \end{subfigure}
        \begin{subfigure}[b]{0.46\textwidth}
        \centering
        \includegraphics[width=\textwidth]{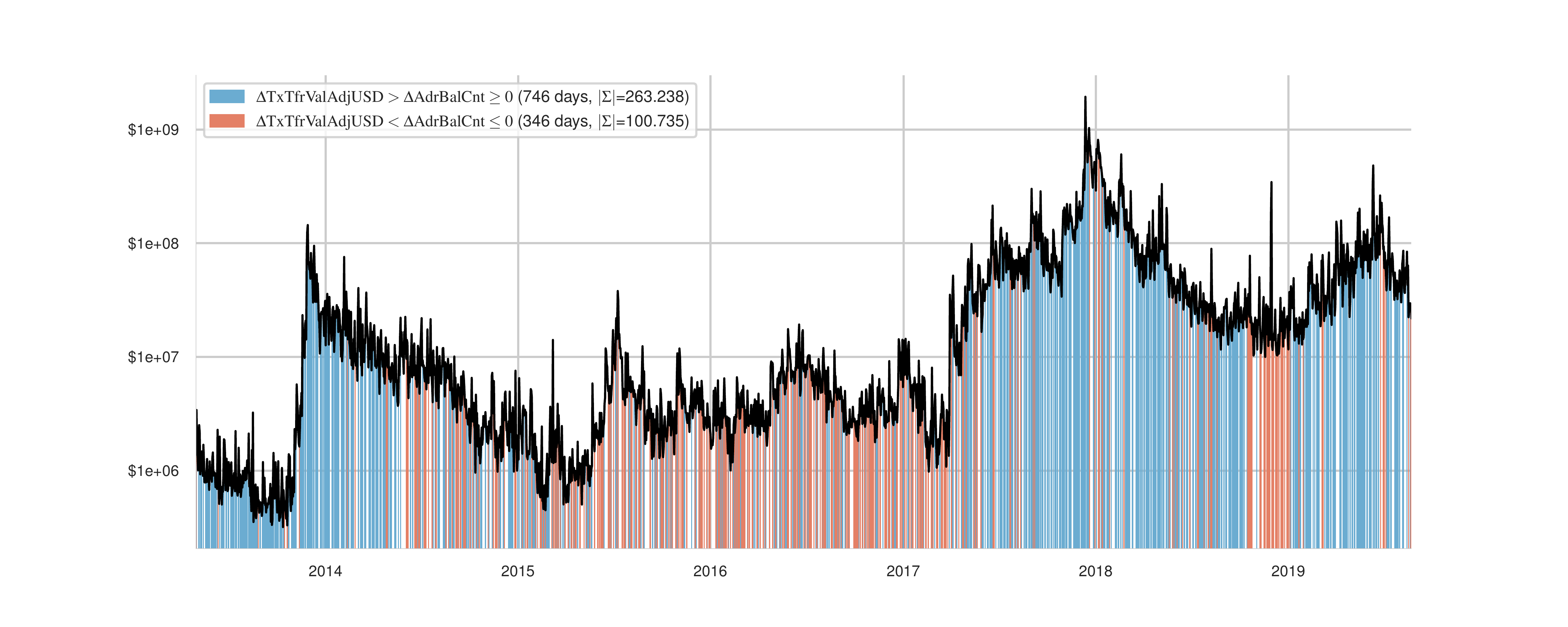}
        \caption{LTC}
         \label{fig:ltc-stem_tx}
    \end{subfigure}
        \begin{subfigure}[b]{0.46\textwidth}
        \centering
        \includegraphics[width=\textwidth]{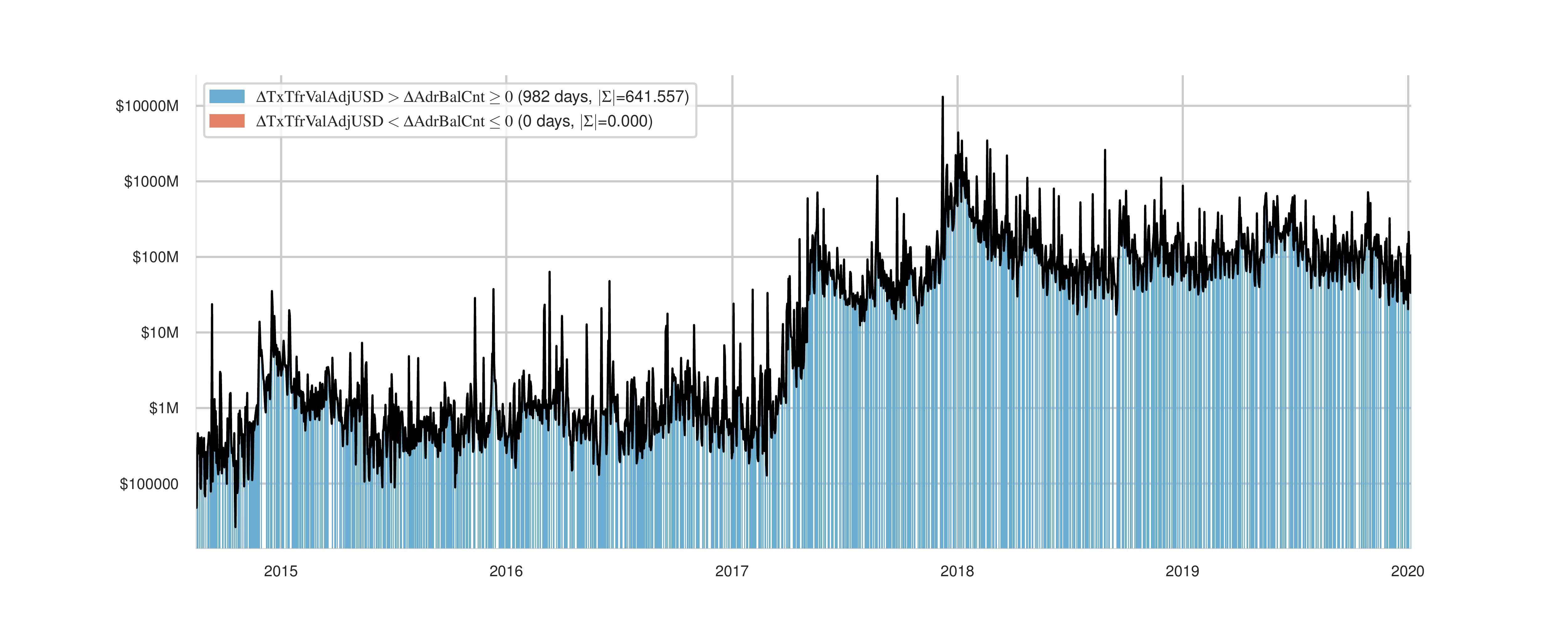}
        \caption{XRP}
         \label{fig:xrp-stem_tx}
    \end{subfigure}
        \begin{subfigure}[b]{0.46\textwidth}
        \centering
        \includegraphics[width=\textwidth]{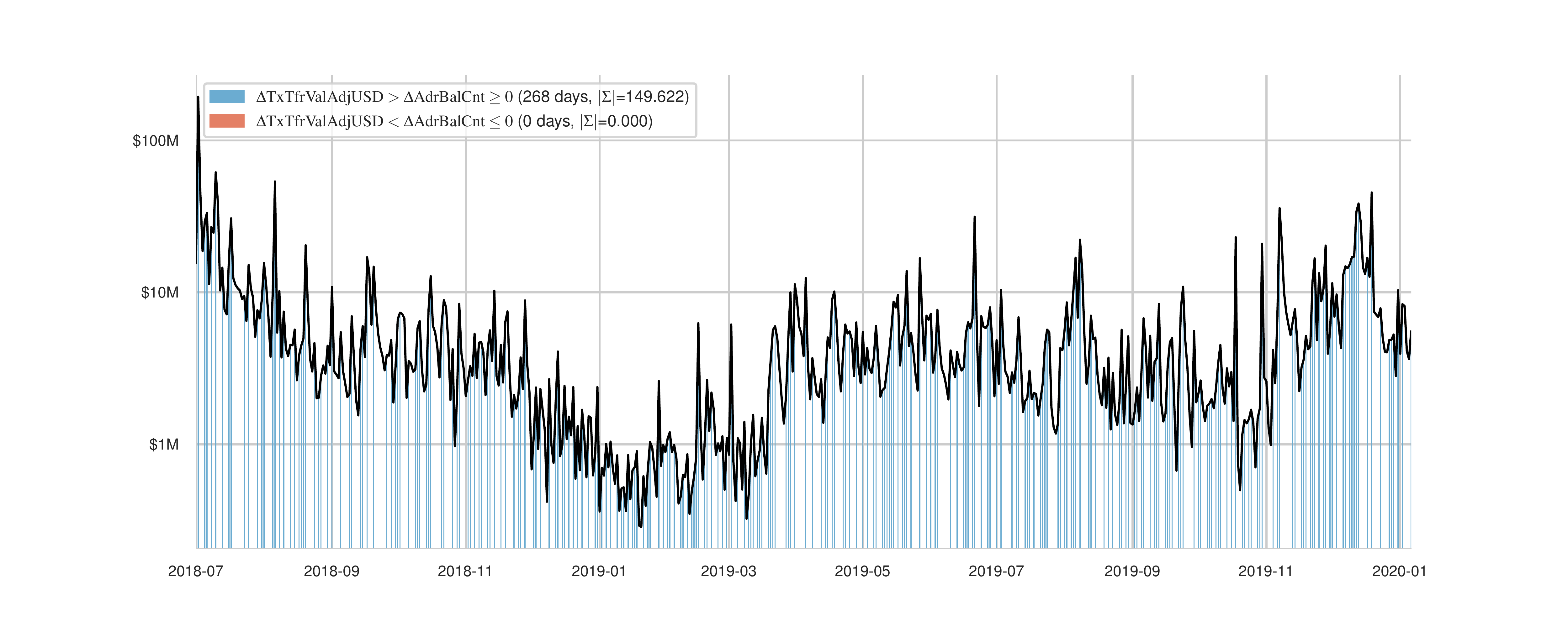}
        \caption{XTZ}
         \label{fig:xtz-stem_tx}
    \end{subfigure}
    \caption{Network effect observations (blue: positive NE, red: reverse NE, white: no NE); userbase measured by total addresses with non-zero balance, value measured by transaction value.}
    \label{fig:stemplots_tx}
\end{figure*}

\paragraph{(5) Network effects are not consistently observed during the early days of cryptoassets and therefore it is doubtful that they can be relied on as a tool to bootstrap a new cryptoasset: } 
A common business model when launching new products or services in digital markets is to exploit network effects to quickly establish a growing foothold. Particularly if the product or service is also the first of its kind to hit the market, network effects can dramatically augment the first mover advantage, everything else equal. Our results indicate that network effects are not consistently observed in the studied cryptoassets during their early days (the first year of data); in particular, DOGE, XTZ and LTC do not exhibit consistent positive network effects neither by token price (PriceUSD) nor by transaction value (TxTfrValAdjUSD) as proxies for value (Figures  ~\ref{fig:stemplots} and ~\ref{fig:stemplots_tx}). The lack of consistency is even more pronounced when userbase is measured by active addresses in the trailing 6-month period, which is an instructive measure here, because it tracks recent user activity which is the driver of early adoption. In Figure  ~\ref{fig:stemplots_6m} only BTC and ETH have a claim to positive early network effects and in ETH they are sparser. This suggests that new cryptoassets cannot necessarily hope that network effects will assist in their initial uptake. It is useful to dispel this hypothesis because investors are looking for patterns in events that may trigger valuation changes (e.g. the hypothesis that cryptoasset value as measured in monetary terms increases once the cryptoasset is listed on a major crypto-exchange). 

\paragraph{(6) Comparison between network effects on price and transaction value reveals sensitivity to price, which can be a competitive disadvantage:} 
Extant literature has relied exclusively on token price as the proxy for network value. Using transaction value too helps us draw useful comparisons. For this, it is most instructive to rely on trailing 6-month active addresses as the proxy for userbase, because this proxy is more responsive to value fluctuations. Then, a comparison between the strength of network effects measured by token price (PriceUSD) and by transaction value (TxTfrValAdjUSD) reveals that some cryptoassets experience greater fluctuations in their transaction value relative to their token price. During upturns, network effects tell us that token price and transaction value increase more than the userbase increases, and during downturns, reverse network effects show the opposite. By comparing the ratios among cryptoassets of the sum of network effects when value is measured by token price and the sum of network effects when value is measured by transaction value one can observe differences in how transaction value is affected among cryptoassets. Specifically, the ratios for BTC, DOGE, ETH and LTC are similar ranging from 0.12 to 0.14 for positive network effects and 0.07 to 0.09 for reverse network effects, whereas XRP's is 0.07 and for XTZ's is 0.06 for positive network effects and 0.04 and 0.03 for reverse network effects (compare sum ratios in Figure ~\ref{fig:stemplots_6m} and Figure ~\ref{fig:stemplots_fx_ 6m}). This means that during periods of positive network effects, XRP's and XTZ's transaction value grows more than their token price grows relative to their userbase, and that during periods of reverse network effects, XRP's and XTZ's transaction value drops more than their token price drops relative to their userbase. This kind of increased volatility may be generally undesirable, but it is particularly problematic during downturns (reverse network effects) because it shows that activity on XRP and XTZ networks is more drastically affected making them more sensitive and less resilient, which is a competitive disadvantage. Our results hold too when we look exclusively at 2017 and 2018 as the years with the most sustained price increase and decrease respectively. 

\section{Conclusion}
Network effects can be among the most common and influential factors shaping market dynamics in industries where products and services are built around networks. It is no wonder that they have been cited as a determinant in how cryptoassets grow in value and compete. Our analysis show that while network effects do characterize cryptoassets, they do not result in the usual concentration and competitive advantage implications usually associated with them. Our work also invites further research to determine the exact scope and conditions under which network effects apply. More precise proxies for userbase and value and accounting for exogenous effects are steps in the right direction.

\bibliographystyle{ACM-Reference-Format}
\bibliography{full_bib}

\section{Appendix}

\begin{figure*}
    \begin{subfigure}[b]{0.46\textwidth}
        \centering
        \includegraphics[width=\textwidth]{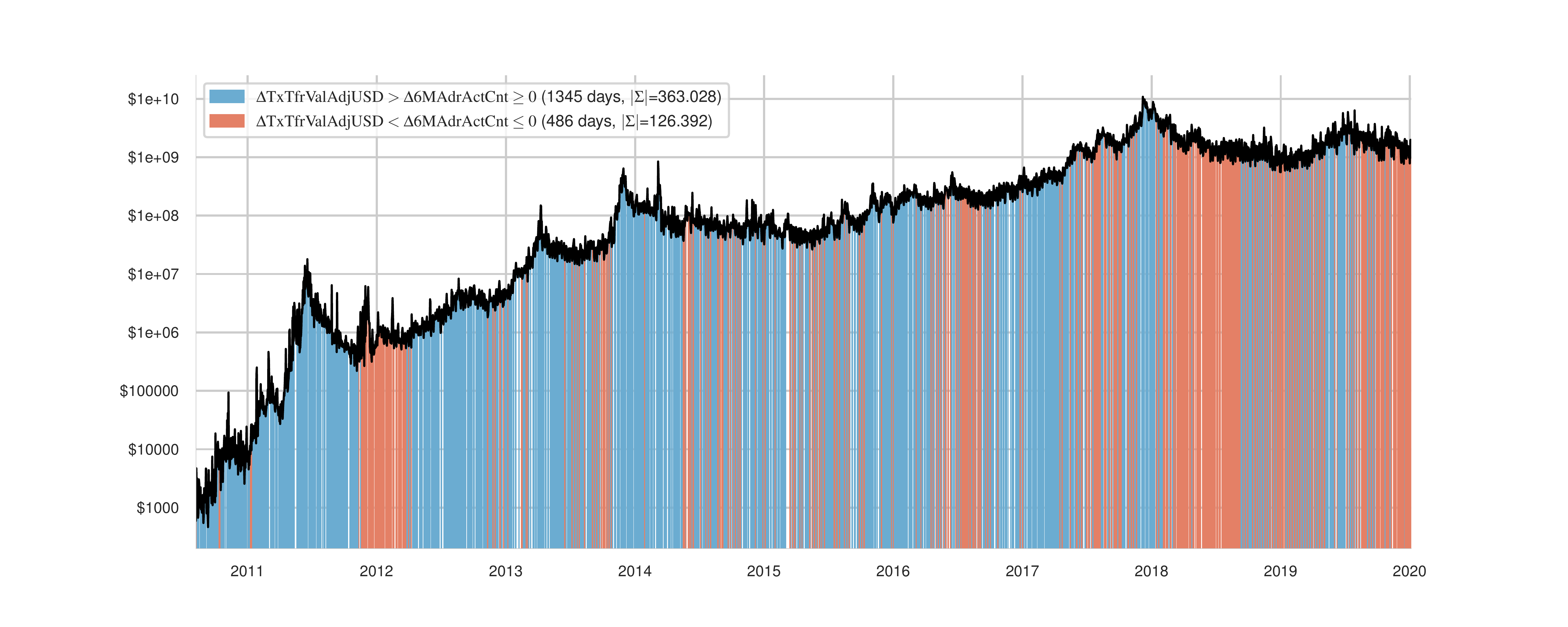}
        \caption{BTC}
        \label{fig:btc-stem_tx_6m}
    \end{subfigure}
    \begin{subfigure}[b]{0.46\textwidth}
        \centering
        \includegraphics[width=\textwidth]{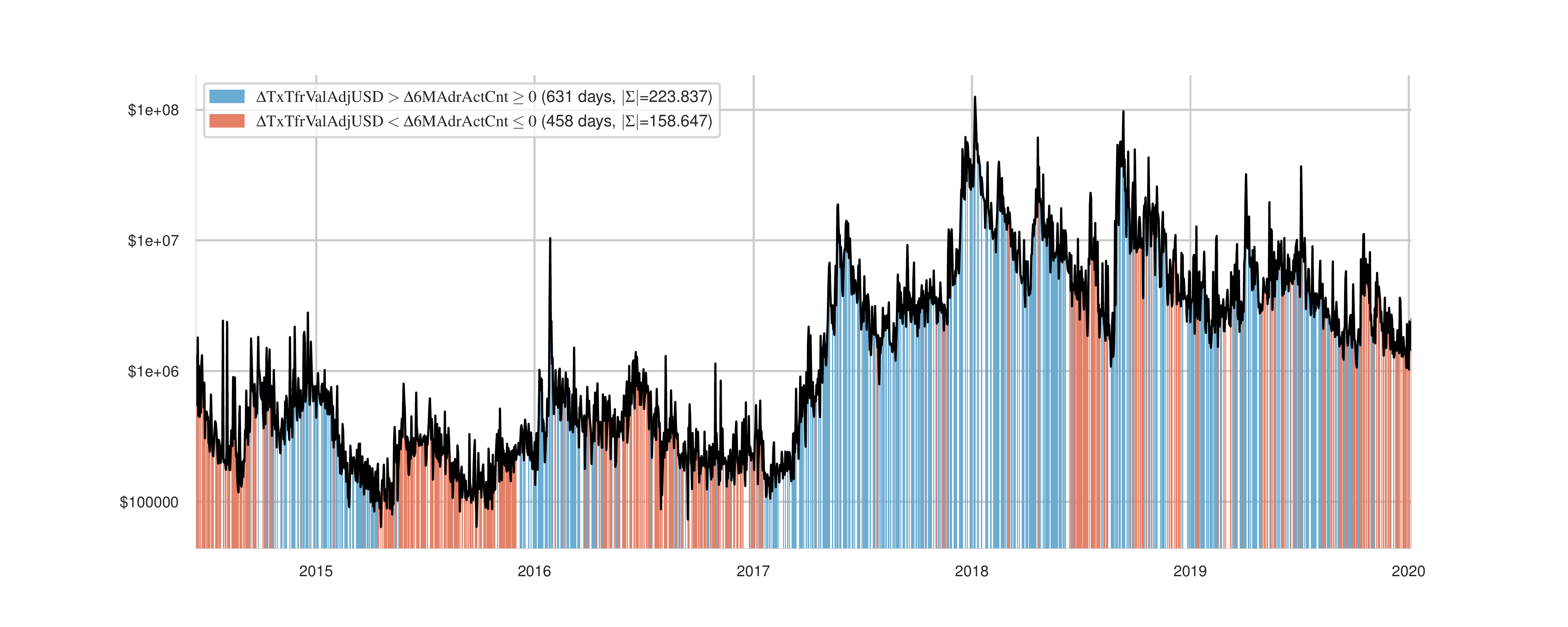}
        \caption{DOGE}
        \label{fig:doge-stem_tx_6m}
    \end{subfigure}
    \begin{subfigure}[b]{0.46\textwidth}
        \centering
        \includegraphics[width=\textwidth]{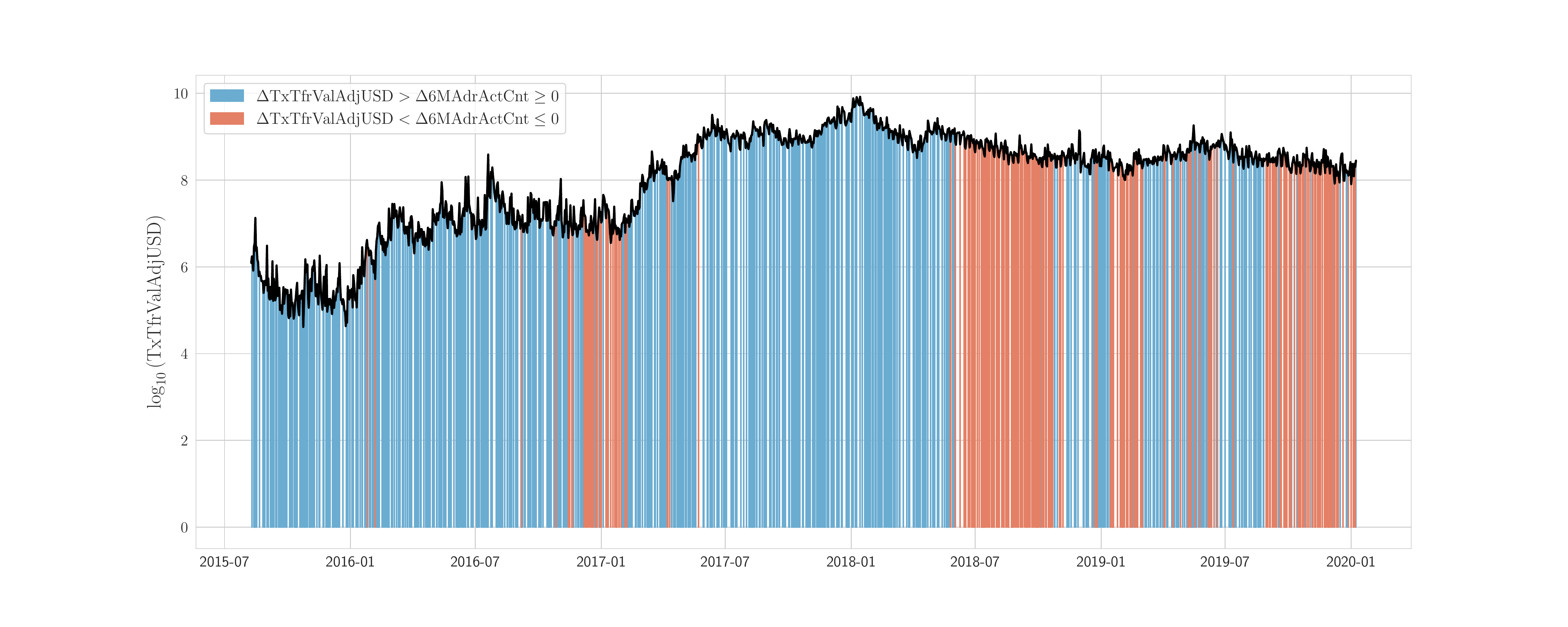}
        \caption{ETH}
         \label{fig:eth-stem_tx_6m}
    \end{subfigure}
        \begin{subfigure}[b]{0.46\textwidth}
        \centering
        \includegraphics[width=\textwidth]{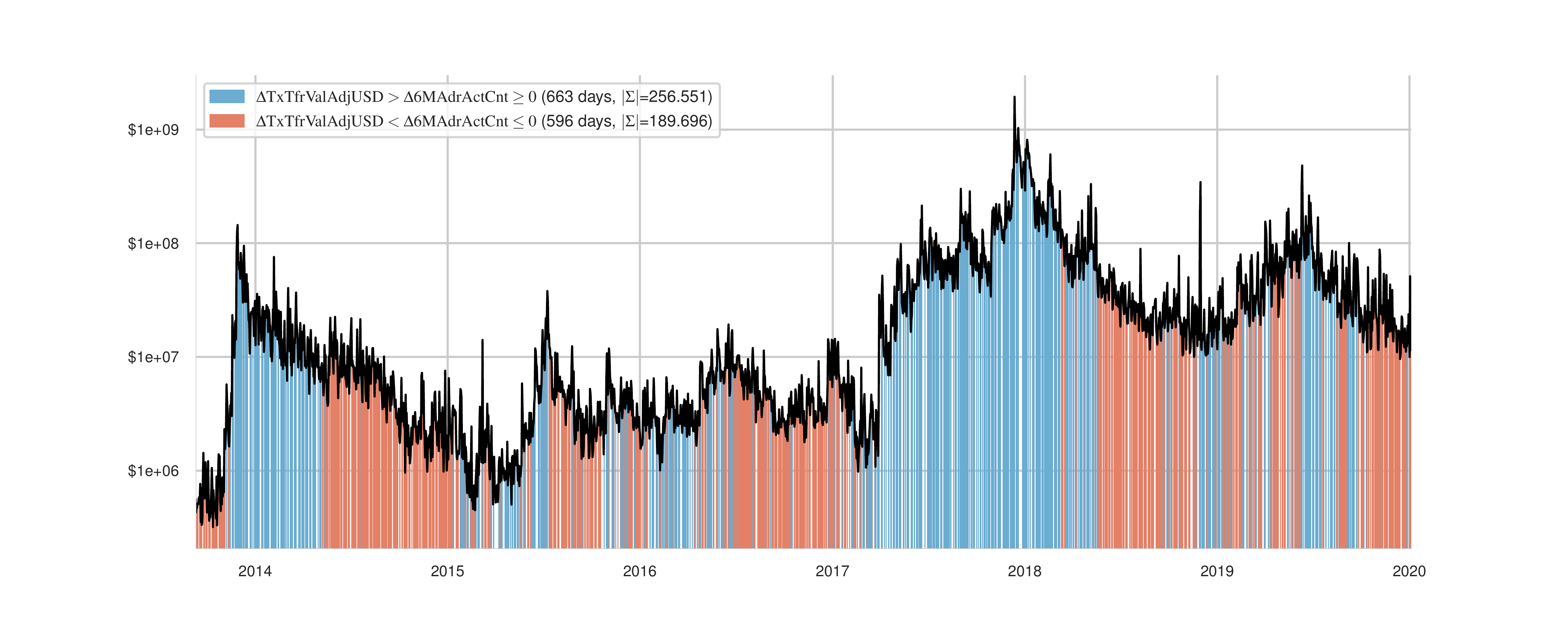}
        \caption{LTC}
         \label{fig:ltc-stem_tx_6m}
    \end{subfigure}
        \begin{subfigure}[b]{0.46\textwidth}
        \centering
        \includegraphics[width=\textwidth]{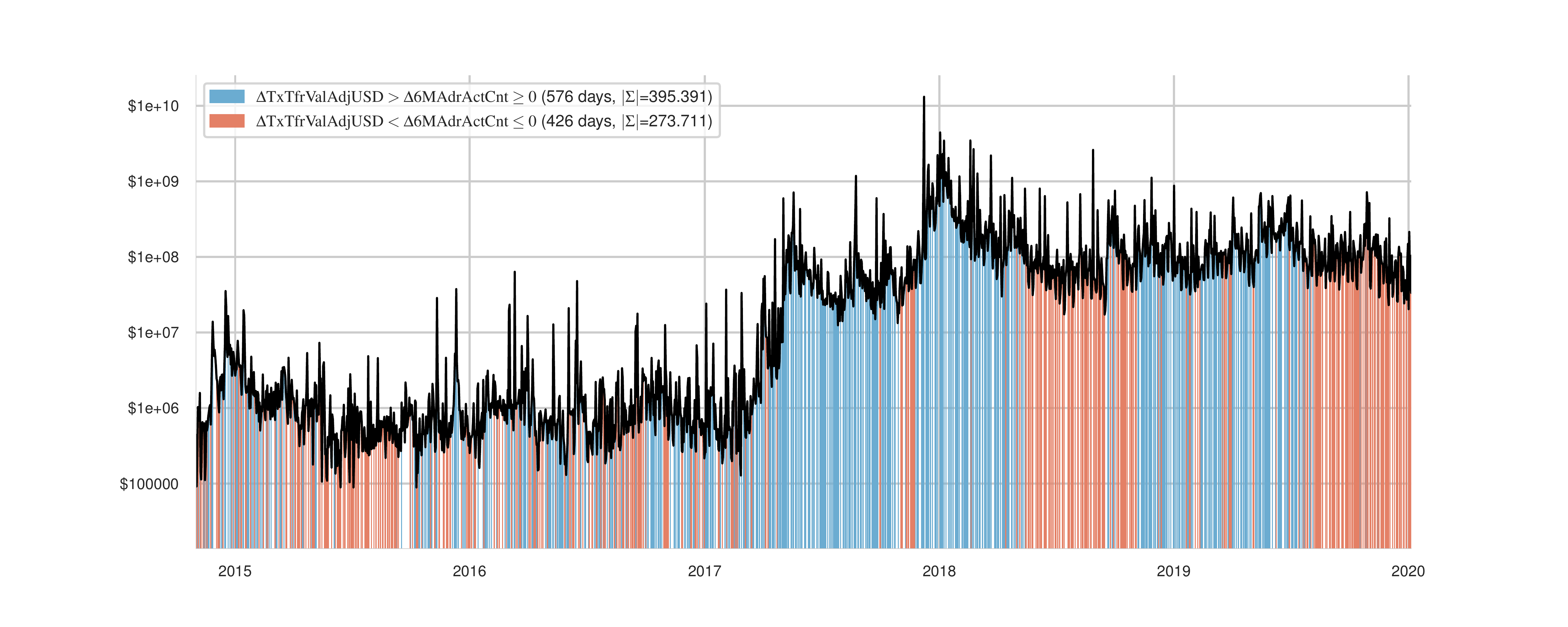}
        \caption{XRP}
         \label{fig:xrp-stem_tx_6m}
    \end{subfigure}
        \begin{subfigure}[b]{0.46\textwidth}
        \centering
        \includegraphics[width=\textwidth]{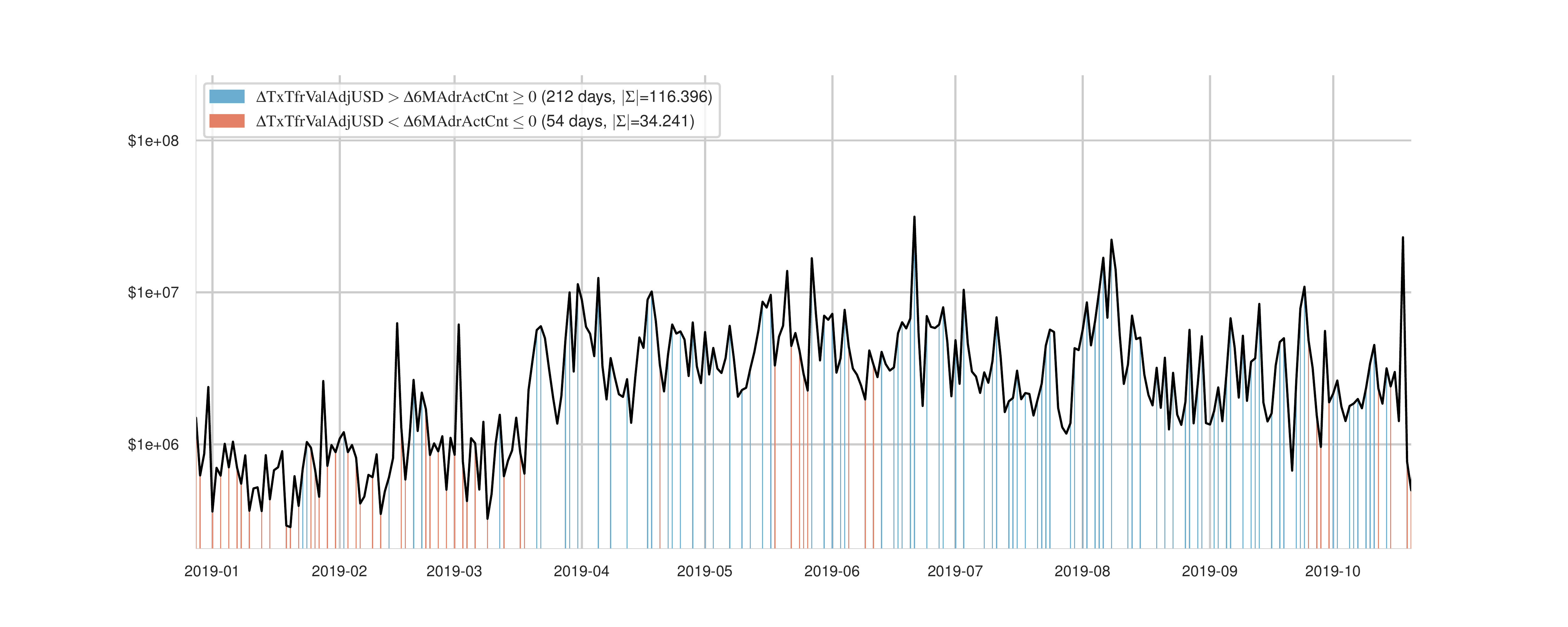}
        \caption{XTZ}
         \label{fig:xtz-stem_tx_6m}
    \end{subfigure}
    \caption{Network effect observations and distribution (blue: positive NFX, red: reverse NFX, white: no NFX); userbase measured by trailing 6 month addresses, value measured by transaction value.}
    \label{fig:stemplots_fx_6m}
\end{figure*}

\end{document}